\documentclass[12pt,preprint]{aastex}

\def\ltsima{$\; \buildrel < \over \sim \;$}
\def\gtsima{$\; \buildrel > \over \sim \;$}
\def\lsim{\lower.5ex\hbox{\ltsima}}
\def\gsim{\lower.5ex\hbox{\gtsima}}
\def\lapp{\ifmmode\stackrel{<}{_{\sim}}\else$\stackrel{<}{_{\sim}}$\fi}
\def\gapp{\ifmmode\stackrel{>}{_{\sim}}\else$\stackrel{<}{_{\sim}}$\fi}

\newdimen\minuswidth    %define @ width of minus sign for tables
\setbox0=\hbox{$-$}
\minuswidth=\wd0
\catcode`@=\active
\def@{\kern\minuswidth}
\setbox0=\hbox{\rm0}
\shorttitle{Double BSS sequence in NGC~362} 
\shortauthors{Dalessandro et al.}
 
\begin{document} 
\title{Double Blue Straggler sequences in GCs: the case of
  NGC~362\footnote{Based on observations collected with the NASA/ESA
    {\it HST}, obtained at the Space Telescope Science Institute,
    which is operated by AURA, Inc., under NASA contract
    NAS5-26555. Also based on WFI observations collected at the
    European Southern Observatory, La Silla, Chile, within the
    observing program 07.D-0188.}  }

\author{
E. Dalessandro\altaffilmark{2},
F. R. Ferraro\altaffilmark{2},
D. Massari\altaffilmark{2}, 
B. Lanzoni\altaffilmark{2},
P. Miocchi\altaffilmark{2},
G. Beccari\altaffilmark{3},
A. Bellini\altaffilmark{4},
A. Sills\altaffilmark{5},
S. Sigurdsson\altaffilmark{6},
A. Mucciarelli\altaffilmark{2},
L. Lovisi\altaffilmark{2}
}
\affil{\altaffilmark{2} Dipartimento di Fisica e Astronomia, Universit\`a degli Studi
di Bologna, Viale Berti Pichat 6/2, I--40127 Bologna, Italy\\
\altaffilmark{3} European Southern Observatory, Karl Schwarzschild Strasse 2, D-85748, Garching bei Munchen,
Germany\\
\altaffilmark{4} Space Telescope Science Institute, 3700 San Martin Drive, Baltimore, MD 21218, USA\\
\altaffilmark{5} Department of Physics and Astronomy, McMaster University, Hamilton, Ontario L8S 4M1, Canada\\
\altaffilmark{6} Department of Physics and Astrophysics, Pennsylvania State University, 525 Davey Laboratory,\\
University Park, Pennsylvania 16802, USA}

\date{09 October, 2013}

\begin{abstract}
We used high-quality images acquired with the WFC3 on board the
{\it HST} to probe the blue straggler star (BSS) population of the Galactic
globular cluster NGC~362. We have found two distinct sequences of BSS: 
this is the second case, after
M~30, where such a feature has been observed. Indeed the BSS location,
their extension in magnitude and color and their radial distribution
within the cluster nicely resemble those observed in M~30, thus
suggesting that the same interpretative scenario can be applied: the
red BSS sub-population is generated by mass transfer binaries, the blue one by
collisions.  The discovery of four new W UMa stars, three of which
lying along the red-BSS sequence, further supports this scenario. 
We also found that
the inner portion of the density profile deviates from a King model
and is well reproduced by either a mild power-law ($\alpha\sim-0.2$) or 
a double King profile.
This
feature supports the hypothesis that the cluster is currently
undergoing the core collapse phase. Moreover, the BSS radial
distribution shows a central peak and monotonically decreases outward
without any evidence of an external rising branch. This evidence
is a further indication of the advanced dynamical age of NGC~362: in fact,
together with M~30, NGC~362 belongs to the family of 
dynamically old clusters (Family III) in the "{\it dynamical
  clock}" classification proposed by Ferraro et al. (2012). The
observational evidence presented here strengthens the
possible connection between the existence of a double BSS sequence
and a quite advanced  dynamical status of the parent cluster.
\end{abstract}

\section{INTRODUCTION}
Among the large variety of exotic objects (like X-ray binaries,
millisecond pulsars, etc.) which populate the dense environment of
Galactic globular clusters (GGCs; see Bailyn 1995, Paresce et
al. 1992, Bellazzini et al. 1995, Ransom et al. 2005, Pooley \& Hut
2006, Freire et al. 2008), Blue Straggler Stars (BSS) surely
represent the most numerous and ubiquitous population.  BSS were
observed for the first time in the outer regions of the Galactic GC M~3
(Sandage 1953).  Since then, they have been detected in any properly
observed stellar system (GGCs, see Piotto et al. 2004, Leigh et
al. 2007; open clusters -- Mathieu \& Geller 2009 -- dwarf galaxies --
Mapelli et al. 2009). BSS are brighter and bluer than the main
sequence turnoff (MSTO), thus mimicking a population significantly
younger than normal cluster stars. Indeed, observations demonstrated
that they have masses larger than that of MSTO stars ($m= 1.2 - 1.7~
M_\odot$; Shara et al. 1997; Gilliland et al. 1998; De Marco et
al. 2004).  However, stellar evolution models predict that single
stars of comparable mass generated at the epoch of the cluster
formation should have already evolved away from the MS; thus some
mechanisms must have been at work to increase the mass of these objects in
their relatively recent past ($2-3$ Gyr ago; Sills et al. 2002).

Two main formation scenarios for BSS have been proposed over the
years: mass transfer (MT-BSS) and direct collision (COL-BSS).  The
collisional formation channel between two single stars was theorized
for the first time by Hills \& Day (1976). Following works (Lombardi
et al. 2002; Fregeau et al. 2004) showed that BSS may form also via
collision between binary-single and binary-binary systems.  In the
mass transfer scenario (McCrea 1964; Zinn \& Searle 1976; Leonard
1996), the primary star transfers material to the secondary one
through the inner Lagrangian point when its Roche Lobe is filled. In
this picture the secondary star becomes a more massive MS star (with a
lifetime increased by a factor of 2 with respect to a normal star of
the same mass -- McCrea 1964) with an envelope rich of gas accreted
from the donor star.  Chemical anomalies are expected for MT-BSS
(Sarna \& De Greve 1996), since the accreted material (currently
settled at the BSS surface) could come from the inner region of the
donor star, where nuclear processing occurred.
%This envelope can show chemical anomalies 
%(a depletion of C and O and an enhancement of N) if the accreted gas comes from 
%regions of the donor stars, which are deep enough to be affected by CNO burning.
% with a hydrogen-rich envelope. Its lifetime along
%the MS will be double than that of a normal star with the same mass. 
%According to the evolutionary phases when the donor fills up its Roche Lobe, 
%MS, SGB/RGB or HB phase, we can schematically divide the mass-transfer 
%scenario in case A, case B and case C respectively (Kippenhahn \& Weigert 1968). \\
Spectroscopic results  supporting the occurrence of the MT formation channel in a few BSS have been recently obtained 
(Ferraro et al. 2006a; Lovisi et al. 2013). Conversely, surface
chemical anomalies are not expected for COL-BSS (Lombardi et al. 1995), since no 
significant mixing should occur
between the inner core and the outer envelope.

In GCs, where the stellar density significantly varies from the center
to the external regions, BSS can be generated by both processes (Fusi
Pecci et al. 1992, Bailyn 1992, Ferraro et al 1995).  Recent works
suggested that MT is the dominant formation mechanism in low density
clusters (Sollima et al. 2008) and possibly also in high-density
clusters (Knigge et al. 2009).  However the discovery of two distinct
sequences of BSS in M~30 (Ferraro et al. 2009, hereafter F09) clearly
separated in color further supports the possibility that both
formation channels can coexist within the cluster core. In fact the
blue BSS sequence is nicely reproduced by collisional models (Sills et
al 2009), while the red one is compatible with binary systems
undergoing MT (see Tian et al. 2006).  The origin of the double
sequence might be possibly related to the core-collapse process that
can trigger the formation of both red and blue BSS, enhancing the
probability of collisions and boosting the mass-transfer process in
relatively close binaries.  Given the evolutionary time-scales for
stars in the BSS mass range, the fact that the two sequences are still
well distinguishable is a clear indication that core collapse occurred
no more than $1-2$ Gyr ago.
%If the proposed picture
%is correct, the double BSS sequence in M~30 allowed for the first time to precisely date, on
%pure observational basis, such a tremendous event of  the dynamical evolution of GCs. \\

Independently of their formation mechanism, BSS are surely the
brightest among the most massive stars within their host cluster, with
a mass that can be even three times larger than the average stellar
mass in old stellar systems ($\langle m\rangle\sim 0.5 M_{\odot}$). For
this reason they are ideal tools to probe the dynamical
evolution of stellar systems.  The radial distribution of BSS with
respect to normal cluster populations (like Horizontal Branch -- HB --
and Red Giant Branch -- RGB) in GGCs is typically found to be bimodal
(see Dalessandro et al. 2008a and references therein): strongly peaked
in the central regions, decreasing at intermediate distances from the
center and rising again in the outskirts. A few exceptions to this
general rule are observed: in $\omega$~Centauri (Ferraro et al. 2006b),
NGC~2419 (Dalessandro et al. 2008b) and Palomar~14 (Beccari et al. 2011)
BSS share the same radial distribution as the normal cluster stars;
instead in a few other cases such as M~80 (Ferraro et al. 1999a; Ferraro et
al. 2012, hereafter F12), M~79 (Lanzoni et al. 2007a), M~75 (Contreras
Ramos et al. 2012) and M~30 (F09, F12), the BSS distribution is monotonic,
with a high peak in the innermost regions and rapidly declining
outward with no signs of rising branch.  Simple Monte Carlo
simulations (Mapelli et al. 2004, 2006) have shown that the observed
radial distribution of BSS is reproduced by assuming that both
formation processes are active, with COL-BSS contributing only to the
central peak, and MT-BSS being necessary to account for the external
rising branch. More recently F12 suggested that the BSS radial
distribution can be used as a powerful indicator of the cluster dynamical age  (the {\it dynamical
  clock}). 
% different stages of
%the clusters dynamical history and they proposed an innovative tool 
%(the {\it dynamical clock}) able to measure the dynamical age of stellar systems from the shape of the BSS
%radial distribution. 
In this picture, GCs with a flat BSS distribution  are quite young
stellar systems, clusters with a centrally peaked and monotonically decreasing BSS distribution
are dynamically old, and those  showing a bimodal distribution have intermediate
dynamical ages (their degree of dynamical aging being a function  of the radial position of
the minimum in the BSS distribution). In this context M~30 belongs to the family of
dynamically old GCs,  in very good agreement with its status of post core-collapsed cluster
suggested by the shape of its density  profile and with the proposed interpretation of its
double BSS sequence (F09).

In order to further explore the link between the presence of a double sequence of BSS and the occurrence of core collapse, 
we acquired {\it Hubble Space telescope (HST)}
images with the Wide Field Camera 3 (WFC3) for a sample of suspected post-core collapse GGCs.  
Here we present the first results of this project and we report on
the discovery of the second case of a BSS double sequence, in NGC~362.
%We find that the BSS population of NGC~362 share very similar properties to that of M~30, both 
%in the optical Color Magnitude Diagram (CMD) 
%and for their spatial distributions. 
%We also report the identification of four new variables 
%along the BSS sequences.  
%This cluster has long been suspected to be a post-core-collapse system (Harris 1996).
%By taking advantage of our high-resolution database, we analyze in detail the star-counts density profile 
%and we compare it with theoretical models.     
The paper is structured as follows: in Section~2 we present the data-base and we describe the photometric 
analysis, as well as the calibration and astrometry procedures adopted. In
Section~3 the determination of the center and density profile is described and a comparison with previous studies
about the dynamical 
state of NGC~362 is performed. In Section~4 we briefly 
discuss the relative proper motion analysis adopted to "clean" the BSS sample from contaminating field stars 
and in Section~5 we go in detail
about the detection of the BSS double sequence and its comparison with M~30. Finally in Section~6 we discuss our
main results.

\section{OBSERVATIONS AND DATA ANALYSIS}

In the present work we used data acquired with the UVIS channel of the 
WFC3 on 2012 April 13
(Proposal ID: 12516; PI: Ferraro).
% This proposal aims at understanding the interplay between the presence 
%of a BSS double sequence and the occurrence of the core collapse in four GGCs.\\
The WFC3/UVIS camera consists of two twin chips, each of $4096\times2051$ pixels, separated by a gap
of approximately 30 pixels. The pixel scale is $0.04 \arcsec$ pixel$^{-1}$, therefore 
the resulting Field of View (FoV) is $162\arcsec \times 162\arcsec$. 
The cluster is approximately centered on chip\# 1 (Figures~1 and 2). 
The dataset is composed of fourteen exposures obtained through the F390W (hereafter U\footnote{Although
the F390W filter almost corresponds to the broad filter C  
of the Washington photometric system (Canterna 1976),
we prefer to label it ``U filter" since it is more popular.}) filter, each one with an exposure 
time $t_{\rm exp}=348$ sec, ten F555W (V) images with  $t_{\rm exp}=150$ sec and fifteen F814W (I) frames with 
$t_{\rm exp}=348$ sec.
Each pointing is dithered by a few pixels in order to allow a 
better subtraction of CCD defects, artifacts and false detections.
For the analysis we used the set of images processed, 
flat-fielded and bias subtracted by standard HST pipelines (\texttt{$\textunderscore$flt} images). 
The data reduction has been performed independently on each exposure by using the publicly available 
point spread function (PSF) fitting software \texttt{img2xym$\textunderscore$wfc3uv}, which is mostly based on
\texttt{img2xym$\textunderscore$WFC} (Anderson et al. 2008). 
Pixel-area effects have been applied to the derived fluxes and geometric distortions 
have been corrected by using the geometric distortion solution provided by Bellini et al. (2011). 

For each photometric band, the obtained star lists have been combined
in order to create a filter master catalog (FMC),
consisting of stars measured in at least five different frames.
For each star, different  magnitude estimates have been homogenized (see Ferraro et al 1991, 1992) 
and their weighted mean and  standard deviation  have been finally adopted as the star magnitude and its photometric error. We then combined the three FMCs to create the final star list, requiring that a star
is present in at least 2 FMCs (i.e. it has two measured magnitudes).
  
%The final catalog counts $\sim 71000$ objects. \\
Instrumental magnitudes have been reported to the VEGAMAG photometric system 
by adopting the zero points reported in the WFC3 web page\footnote{http://www.stsci.edu/hst/wfc3/phot$\textunderscore$zp$\textunderscore$lbn} 
for a $0.4\arcsec$ aperture correction. Stars with I$<16.5$ suffer from non-linear CCD response and saturation problems;
thus they were not considered for the BSS analysis.
The analysis of the BSS double sequence has been performed by using exclusively WFC3 data (see Section~5.1)
Instead additional 
HST and ground-based observations have been considered
for the determination of the cluster center, to build the star-count density profile all over the entire radial extension of the cluster and to
study the BSS radial distribution. 
In particular, we used images obtained with the Wide Field Imager (WFI) mounted at the MPG/ESO 2.2m telescope.
The WFI frames consists of eight $4096\times2048$ pixel chips with a spatial resolution 
of $ 0.228\arcsec$ pixel$^{-1}$, thus each WFI exposure covers a FoV of about $33'\times34'$.\\
We used 5 long exposures (Prop: 07.D-0188(A); PI: Ortolani), two obtained through the B band with 
$t_{\rm exp}=360$ sec each, and 
three in V band with $t_{\rm exp}=300$ sec. We also used two short exposures, one B and one V, 
with $t_{\rm exp}=20$ sec each, to properly measure the bright portion of the RGB and Asymptotic Giant
Branch sequences.  
Master bias and flat-fields have been obtained by using a large number of calibration
frames. Scientific images have been corrected for bias and flat-field by using standard procedures and tasks  
contained in the Image Reduction and Analysis Facility (IRAF)\footnote{IRAF is distributed by the National Optical
 Astronomy Observatory,
which is operated by the Association of Universities fro Research in Astronomy, Inc., under the cooperative 
agreement with the National Science Foundation.}.
The photometric analysis has been performed on each chip separately by using \texttt{DAOPHOT{\sc II}} (Stetson 1987).
For each frame we selected several tens of bright and relatively isolated stars to model the PSF.
We used a Moffat analytic function and a first order spatial variation for the PSF. 
For each chip we obtained a star list that was then combined by using \texttt{DAOMATCH} and \texttt{DAOMASTER}.
Only stars measured at least twice in each band were considered.
We used the B and V magnitudes of the Stetson photometric secondary standard catalog (Stetson 2000) to report our 
instrumental magnitudes B$_{\rm ins}$ and V$_{\rm ins}$ to the standard Johnson photometric system. 
In particular, we analyzed the distributions in the ($B-B_{\rm ins},B_{\rm ins}-V_{\rm ins}$) and 
($V-V_{\rm ins},B_{\rm ins}-V_{\rm ins}$) planes
for the stars in common, and we used their best-fit relations to calibrate the instrumental magnitudes.
We adopted as calibration relations the linear best fits to the two distributions. 
The WFI data have been used to obtain the density profile in the external regions (see Section~3) and
the BSS radial distribution in the area not covered by the WFC3 data-set. (Section~5.3).\\
In the innermost, highly-crowded regions we exploited also the high spatial resolution of the Advanced Camera for Survey (ACS)
High Resolution Channel (HRC). The HRC is a $1024\times1024$ pixels array with a pixel scale 
of $\sim 0.026\arcsec/$pixels, thus covering a FoV of about $28\arcsec\times25\arcsec$. The HRC data-set (Prop ID=10401; PI: Chandar) consists of eleven 
exposures of 85 sec each obtained with the broadband filter F435W ($\sim$ B). 
As for the WFC3 data-set, we used 
{\rm $\textunderscore$flt} images pre-processed by the standard HST pipelines.
Each exposure has been corrected for pixel area effect, by using the most-updated 
Pixel Area Map images available at the HST web site. The analysis has been performed in each frame independently 
by using \texttt{DAOPHOTII} and by selecting some tens of stars to model the PSF. Also in this case 
we adopted a Moffat
analytic PSF and a first order variation. Star positions and magnitudes have been then combined with 
\texttt{DAOMATCH} and \texttt{DAOMASTER}. Instrumental magnitudes have been reported to the VEGAMAG photometric system 
by using the prescriptions and zero points listed in the "ACS Calibration and Zeropoints" 
web page\footnote{http://www.stsci.edu/hst/acs/analysis/zeropoints}. The HRC data have been crucial for an accurate determination 
of the cluster center and to provide a complete sample of stars to built the star-count density profile in the innermost
few arcsec (Section~3).
 
The instrumental positions of stars in each WFI chip have been separately roto-translated
to the absolute ($\alpha$, $\delta$) coordinates by using
the stars in common with the astrometric standards listed in the Guide Source Catalogue 2.3 (GSC2.3) 
and the cross-correlation software {\rm  CataXcorr}.
For each chip we found several tens of stars in common (hundreds in the chip containing the cluster core) 
thus allowing a very accurate 
roto-translation solution. Once all the chips were put on the absolute coordinate system, we merged 
them in a single homogeneous source catalogue.
We then used the stars detected with WFI and falling in the WFC3 FoV as secondary astrometric standards.
In this case several thousands of stars have been found in common. 
Similarly, positions of stars detected with HRC were first corrected for geometric distortions
and then put on the absolute reference system by using the stars in common with the WFC3 catalog.

At the end of the procedure we have three independent star catalogues on the same coordinate reference system. 
As anticipated they have been all used for 
building the density profile (see Section~3) and to determine the center of gravity, 
while we used only the WFC3 star list to study the BSS population.

\section{Determination of the cluster center and density profile}

The dynamical state of NGC~362 is quite uncertain. 
A detailed study has been performed by Fischer et al. (1993), who 
combined photometric data and radial velocities for $\sim200$ RGB stars.
By using single- and multi-mass King-Michie (KM) models  they were able to reproduce well
the surface brightness profile (SBP) of this system. However the observational constraints from the kinematics and rotation 
obtained with the spectroscopic data were not compatible with the best-fit KM models 
in the assumption of isotropy.
In order to reconcile the kinematical measures with the observed SBP
a shallow mass function and intermediate amounts of anisotropy in the velocity-dispersion tensor 
had to be adopted.
On the basis of SBP fitting, this cluster has been classified as a possible 
PCC by Trager et al. (1995; their best fit structural parameters are $c=1.94$ and $r_c=10.2\arcsec$).
Also Harris (1996, version 2010)
classified this system as possible 
PCC ($c=1.76$, $r_c=10.8\arcsec$), while McLaughlin \& van der Marel (2005)
were able to reproduce its SBP with a 
King model with parameters $c=1.80$, $r_c=10.1\arcsec$ without discussing the
possibility that this system already underwent the collapse of the core.

In order to clarify the complex picture of the dynamical state of NGC~362, we 
used the entire data-set described in Section~2 to obtain its surface density profile from resolved star counts.
First we determined the center ($C_{\rm grav}$) of NGC~362,
by averaging the positions  $\alpha$ and $\delta$ of selected stars lying in the HRC FoV 
(see for example Ferraro et al. 2003a; Dalessandro et al. 2009).
We preferred to use the HRC data because of the many saturated 
stars in the inner regions of the cluster present in WFC3 images, and because of the higher
resolution of the HRC camera. This choice however limits the distance from the center within which we 
can average the positions of stars.
We chose the center listed by Goldsbury et al. (2010) as starting guess point of our 
iterative procedure (Montegriffo et al. 1995).
In order to avoid spurious or incompleteness effects, 
we performed different measures averaging the position of stars selected in different 
magnitude intervals\footnote{Note that for the HRC sample, the limits in V have been inferred by using the stars 
in common with WFC3 catalog (see leftmost panel of Figure~3).} (all within the range $12.5<$V$<18.5$) and distance ($d$)
from the 
starting guess center ($8\arcsec<d<12\arcsec$).
The resulting  $C_{grav}$ is the average of
these measures and it is located at $\alpha_{J2000}=01^h:03^m:14^s.099$, $\delta_{J2000}=-70^{\circ}:50\arcmin:56\arcsec .04$ 
($RA=15.8087453$, $Dec=-70.8489012$). The estimated $C_{\rm grav}$ turns out to be at a distance $d\sim0.90\arcsec$ 
(corresponding to $\sim 1\sigma$ of our measures) 
in the South-West direction from the one listed by Goldsbury et al. (2010). 
%As a sanity check we also computed $C_{grav}$ as the "density center" (see for details Casertano \& Hut 1985).
%In brief, for a given radius centered on a starting guess point and for the stars located within it, 
%the local density around the position of the {\it i}th star is evaluated as the inverse of its squared 
%distance from the sixth nearest star. Then the cluster density is computed as the density weighted average 
%of the N star positions. The evaluation is repeated by using the last found center until the difference between 
%two consecutive measures is smaller the $0.01\arcsec$. The values thus obtained agrees with the value
%quoted above within the uncertainties at $1\sigma$ level. 

By properly combining the HRC, WFC3 and WFI data-sets, we constructed the density profile 
by using direct star counts along the entire radial extension of the cluster.
As shown in Figure~3, we typically used stars in the magnitude range $17.5<$V$<19.5$ 
in order to avoid incompleteness and saturated stars. 
When possible, however, as in the case of the HRC sample, which does not show 
saturation problems, we extended our selection to stars brighter than this limit.
We sub-divided the total FoV in 30 concentric annuli centered on $C_{\rm grav}$ and reaching $r=1300\arcsec$.
For $r<16\arcsec$ we used only stars in the HRC FoV, for $16\arcsec<r<100\arcsec$ stars in the WFC3 FoV and for 
the outer regions we used the WFI catalog. 
Each annulus was then split in an adequate number of sub-sectors (2 or 4 depending on the number of stars).
In each sub-sector the density has been estimated as the ratio between the
selected number of stars counted and the covered area. The density assigned to a given annulus 
is the average of the densities of each
sub-sector of that annulus. Densities thus obtained have been corrected for incomplete area 
coverage due to the WFI gaps.  The error assigned to each density measure is defined as 
%the quadratic sum
%of the pure Poissonian error ($\sqrt{N}$) and 
the dispersion from the mean of the sub-sector densities. Overlapping annuli were used to check and apply normalizations 
between the three subsamples. The resulting density profile
is shown in Figure~4 (open squares).
As clearly visible in the right panel of Figure~3, NGC~362 is contaminated by fore- and back-ground Galaxy stars and even more strongly by
stars belonging to the Small Magellanic Clouds (SMC; see the population at $(B-V)\sim0$ and $(B-V)>0.6$, respectively). 
We estimated the background density by using the six outermost points 
(with distances from $C_{grav}$ $r>450\arcsec$), that have almost the same density,  
describing a sort of plateau
\footnote{For the most distant annulus ($1000\arcsec<r<1300\arcsec$), variations are observed as a function 
of the azimuthal angle in which the density is calculated. In particular the density is slightly larger toward the SMC direction 
(even if this discrepancy has not a high statistical significance).
However this effect has a negligible impact on the overall background determination.}.
%We note however that the most distant point ($1000"<r<1300$) has a slightly larger value than the others 
%(even if this discrepancy has not a high statistical significance).
%The amplitude of this difference is a function of the azimuthal region in which the density is calculated.
%In particular we observe that this effect is larger in the SMC direction, thus it is likely due to the 
%fact that the SMC stars density is not constant within the WFI FoV.
We subtracted the mean background value to all the other annuli and the resulting density profile is shown
in Figure~4 (black filled circles). As expected, after the background correction, the profile changes sensibly in the outermost regions, 
while it remains unchanged in the innermost ones.

We performed a fit of the density profile by using a single-mass King model (King 1966). 
We excluded from the fit the three innermost points ($r<3\arcsec$) of the observed profile, 
since they appear to deviate from a flat core behavior (see the inset in Figure~4). 
The best fit model is obtained
for a core radius $r_c=13\arcsec$, concentration $c=1.74$ and limiting radius $r_t\sim720\arcsec$.
The parameters obtained from our analysis are slightly 
different from those obtained in previous works (in particular, $r_c$ is about $30\%$ larger, while $c$ is up to 
$15\%$ smaller).
However our estimates cannot be directly compared with previous determinations 
because they included also the innermost $3\arcsec$ (which are instead
excluded in our analysis).
The innermost portion of the observed density profile ($r<3\arcsec$) is well fitted by a 
mild power-law with slope $\alpha\sim-0.2$.
The derived  power-law is shallower than what typically observed (and expected) for post core-collapse
clusters 
(for example $\alpha\sim-0.5$ for M~30; F09). 
It is worth noting, however, that N-body simulations (Vesperini \& Trenti 2010) show that the presence of
relatively shallow cusps in the density and SBPs could be found in clusters
during the phase of either pre-core-collapse or core-collapse. On this basis we can argue that NGC~362 
is possibly experiencing the core-collapse event.\\
We also tried to fit the innermost portion of the profile with an additional King model, as done by Ferraro et al. (2003b)
for the case of NGC~6752. In particular we found that for $r<3\arcsec$ the profile is well fit by a King model with 
$c=2.9$ and $r_c=4.7\arcsec$ (see inset in Figure~4).
Also this behavior would suggest that NGC~362 is a post core-collapse cluster experiencing a post-core-collapse bounce, 
i.e. a large amplitude oscillation in the core due to gravothermal instability of collisional systems (Cohn et al. 1991).   
We also tried to reproduce the entire surface density profile with
multi-mass King models, but the fit always resulted of significantly lower quality with respect to  
the best solution described above.\footnote{A mono-mass King model with the same structural parameters but including
a central IMBH (Miocchi 2007)
with $M_{IMBH}/M_{cluster}=3\times10^{-3}$ is able to reproduce the innermost $3\arcsec$ of the observed profile better than the
multi-mass models.}

\section{Proper motion analysis}

As evident in Figure~5 and as already discussed in previous Section 3,
the CMD of NGC~362 is contaminated by fore and background Galaxy stars and even 
more strongly by the SMC populations. In particular the MS of the SMC defines a quite clear sequence
at $V>20$ and $0<(U-I)<2$. 
%Such contamination would prevent us from any accurate conclusion about the BSS population,
%thus we performed a relative proper motions analysis 
%in order to discriminate between member and non member stars in 
%the WFC3 FoV and to build a clean catalogue of stars properly selected on the basis of their membership.\\
In order to evaluate the level of contamination in the direction of the cluster, we performed a relative 
proper motion analysis.
For this purpose we complemented our HST WFC3 images with a first epoch data-set consisting of Wide Field Camera ACS images 
obtained in June 2006 (Prop ID 10775; PI: Sarajedini). The proper motion analysis has been performed 
following the approach described in Anderson et al. (2010).
Briefly, the procedure consists in measuring the 
displacement of the instrumental (x, y) coordinates between the positions of stars in
the first epoch and the corresponding positions in the  second one, once a common reference frame is defined.
The first step is to adopt a distortion-free reference frame. The ideal choice was to adopt as reference frame
the one published by Anderson et al. (2008).
%which has been corrected for
%geometric distortion following prescriptions by Anderson (2005).
The second step is to find accurate transformations between each single-frame
catalogue and the reference frame. In order to do that, we strictly selected 
unsaturated stars with magnitude ranging between $18<mag<21$ in each band.
Moreover, we selected only stars distributed along the MS and the sub-giant branch (SGB) of NGC~362,
in order to derive global 6-parameters linear transformations by using stars with a high probability to be cluster members.
For this selected sample of stars (hereafter the reference stars) the residuals of the transformations are always smaller
than $0.03$ pixels (which accounts for measurement errors and stars' internal motion).
We then applied the derived transformations to all the stars detected in each frame.
Every single-frame catalogue coordinate (x, y) has been transformed
onto the distortion free reference exposure. We will refer to this transformed
coordinate as ($x_{t}$, $y_{t}$). The mean position of a single star
in each epoch ($x_m$, $y_m$) has been measured as the 3-sigma clipped mean position calculated 
among all the N individual single-frame measurement ($x_{t}$, $y_{t}$), and the relative 
r.m.s. of the position residuals around the mean value divided by $\sqrt(N)$ has been used as associated error ($\sigma$). 
Finally, displacements are obtained as the difference of the positions ($x_m$, $y_m$) between the two epochs
for all the stars in common. 
The error associated to the displacement 
is the combination of the errors on the positions of the two epochs.
We iteratively repeated the procedure, by rejecting from the initial list of reference cluster stars those whose motion was not
consistent with the cluster mean motion, and we re-calculated new, improved linear transformations. Thus, according to the procedure described above,
only stars with residuals ($\sigma_X=|x_t-x_m|$ and $\sigma_Y=|y_t-y_m|$) smaller than 0.8 pixels were considered after
each iteration. The convergence is assumed when the number of reference stars that undergoes this
selection changes less than the $5\%$ between two subsequent steps. 
The displacements
obtained with this approach are converted in relative proper motions, measured in pixel yr$^{-1}$ (here the pixel is that of
the reference frame $0.05\arcsec$ pixel$^{-1}$), by dividing for the temporal baseline ($\sim5$ yr).

The upper panels of Figure~5 present the vector point diagram (VPD), where we can distinguish 
two main sub-populations.
The first and dominant one, centered at ($\mu_X=0$ pixel yr$^{-1}$, $\mu_Y=0$ pixel yr$^{-1}$) is, 
by construction, the cluster population. 
The second, at ($\mu_X=0.7$ pixel yr$^{-1}$, $\mu_Y=0.2$ pixel yr$^{-1}$) is instead populated by the SMC stars. The separation between the two 
components appears clearly in the (V,U-I) CMDs as shown in the lower panels of Figure~5.
In Figure~6, we show the VPD at different magnitude levels. As expected, 
the population belonging to the SMC starts to appear in the VPD diagram for $V>20$. In addition the distribution of NGC~362
and SMC gets broader as a function of increasing magnitudes, because of the increasing uncertainties on 
the centroid positions of faint stars.
The same behavior is visible at bright magnitudes ($V<17$) because of non-linearity and saturation problems.

To build a clean sample of stars with a high membership probability, we defined in the VPD and for each magnitude bin 
a different fiducial region centered on ($\mu_X=0$ pixel yr$^{-1}$, $\mu_Y=0$ pixel yr$^{-1}$). The fiducial regions have 
radii of $2\times \sigma$, where $\sigma$ is
the dispersion of fiducial member stars, i.e. those with a distance $r<0.3$ pixel yr$^{-1}$  
from ($\mu_X=0$ pixel yr$^{-1}$, $\mu_Y=0$ pixel yr$^{-1}$) (see the member selection in the upper panels of Figure~5).
It is worth noting that 
slightly different criteria for the membership selection do not appreciably affect the results about the BSS population.

\section{The BSS population}
\subsection{The BSS double sequence}

The evidence shown in Section~3 suggests that NGC~362 could have
started the core collapse process.  This makes it particularly
interesting in the context of the working hypothesis proposed by F09
that clusters undergoing the core-collapse process could develop a
double BSS sequence.  For this reason, the BSS population of NGC~362
has been analyzed first in the (V,V-I) CMD, in order to perform a
direct comparison with the observations of M~30 (F09).  Following the
approach adopted in previous papers (Lanzoni et al. 2007b; Dalessandro
et al. 2008b) we selected BSS candidates by defining a box which
roughly selects stars brighter and bluer than the TO point,
corresponding to $V_{TO}\sim19$ and $(V-I)_{\rm TO}\sim 0.7$. As
usual, in our selection we tried to avoid possible contamination of
blends from the SGB, TO region, and the saturation limit of the deep
exposures (dashed line in Figure 6).  With these limits 65 candidates
BSS have been identified.  We emphasize that the selection criteria
are not a critical issue here, since the inclusion or exclusion of a
few stars does not affect in any way the results of the paper.  The
selected BSS are shown in the zoomed region of the CMD in Figure~7. At
a close look it is possible to distinguish two almost parallel and
similarly populated sequences, separated by about $\Delta V\sim0.4$
and $\Delta (V-I)\sim0.15$.  Such a feature resembles the one observed
in M~30.
%that makes their distribution in color and magnitude similar to the double BSS sequence observed in M~30.\\ 

In order to perform a more direct comparison with  M~30, we over-plotted 
two fiducial areas (grey regions in Figure~7) representative of the color 
and magnitude distribution of the red
and blue BSS populations  in that cluster.
Differences in distance moduli and reddening have been properly taken into account.
Figure 7 shows that the BSS of NGC~362 nicely fall within the fiducial regions and only sparsely populate the region between the two. 
Moreover the BSS population of NGC~362 show  luminosity and color extensions  similar to the 
ones observed in M~30.  The red sequence appears slightly more scattered than that observed in M~30. This is mainly due to the fact that in the case of NGC~362 a few candidate BSS at 
$(V-I)\sim 0.6$ have been included
into the sample. However,
we
can safely conclude that, within the photometric uncertainties,
the two BSS sequences of NGC~362 well resemble the red and blue BSS sequences of M~30.
%According to their average color (temperature) and as done by F09, 
%we flagged the two sequences as blue BSS and red BSS  (Figure~7).
The blue BSS sample counts 30 stars which are distributed along a narrow 
and well defined sequence in the (V, V-I) CMD, while the red sequence counts 35 stars.
The relative sizes of the two populations is very similar to what found in M~30.
%In contrast, the red BSS defines a broader sequence with a spread in color . 

We also analyzed the distribution in color and magnitude of the red and blue sequences of both clusters by comparing
their geometrical distance ($d$) from  the same straight line fitting the blue sequence of M~30 used by F09.
To do so, we anchored the two (V, V-I) CMDs at the TO level of M~30 by correcting for relative differences of
reddening and distance moduli.
The result is shown in the upper panel of Figure~8 as an histogram. It reveals the presence of two distinct peaks 
 nicely overlapping with those observed in M~30, 
with a separation $\Delta d\sim0.12$. However, both the blue BSS and the red BSS  distributions of NGC~362 are broader than those
in M~30; 
in particular the red sequence is more spread out by $\sim0.05-0.1$ mag with respect to the one studied by F09.
The bimodality remains clearly visible in the cumulative histogram shown in the lower panel of Figure~8.
In order to quantify the significance of the bi-modality of the distribution shown in Figure~8, we 
used the Gaussian mixture modeling algorithm presented by Muratov \& Gnedin (2010). 
This algorithm evaluates whether a bimodal fit is an improvement over a unimodal one by performing a parametric bootstrap and
using three different statistics:
the separations of the means relative to their widths as defined by Ashman et la (1994), the {\it kurtosis} of the distribution,
and the likelihood ratio test (Wolfe 1971). We obtain that a unimodal distribution is rejected with a $99.6\%$ probability; 
hence the distribution shown in Figure~8 is bimodal with a confidence level of $3\sigma$.
    
We analyzed the spatial distribution of the red and blue BSS populations.
Their relative location in the WFC3 FoV is shown in Figure~2. From this map it is already possible
to make some qualitative considerations: 1) red BSS  appear more centrally concentrated 
than blue BSS, 2) almost the entire BSS population lies in chip\#1.   
We analyzed in further detail this fact by looking at the cumulative radial distribution.
We used  SGB stars in the magnitude interval $18<$V$<18.5$ as reference population. 
Both the red and the blue BSS samples are more centrally concentrated than the reference population. 
A Kolmogorov-Smirnov test  gives a probability 
$P\sim10^{-5}$ that they are extracted from the same parent population. Moreover, red BSS  are
more centrally segregated than the blue ones, with a high ($3\sigma$) confidence level that 
they are extracted from different populations. 
In addition, in striking agreement with what found by F09 in M~30, 
we do not observe any blue BSS within $5\arcsec-6\arcsec$ from $C_{grav}$ and both the blue and
red samples completely disappear (within the WFC3 FoV) at distances $r>75\arcsec$. 
The observational evidence collected so far leads us to conclude that NGC~362 is the second cluster,
after M~30, showing a clear double sequence of BSS.

\subsection{Variable stars in the BSS sample}

The work by Szekely et al. (2007) has strongly contributed to the study of variable stars 
in NGC~362. They have increased by three times the number of known variables in this GC.
They found that NGC~362 hosts a relatively large number of variable stars.
In particular they counted 45 RR~Lyrae stars and 21 other short-period variables 
(like $\delta$~Scuti, eclipsing variables, etc.). 

With the aim of identifying variables in the selected BSS population, and in order to avoid contamination
from back- and fore-ground variables, in particular 
from the many Cepheids and long-period variables belonging to SMC and lying in the BSS region
(see Figure 13 in Szekely et al. 2007),
we cross-correlated the publicly available list of non-RR~Lyrae short-period variables with
our proper motion-cleaned-catalog (see Section~4). 
In our FoV we find two stars in common. Both are SX-Phoenicis, as expected for 
stars crossing the instability strip at the BSS magnitude level.
In the list of Szekely et al. (2007) they are named V52 and V63 and they have no period determination. In particular V63 
exhibits complex multi-periodic variations. Both V52 and V63 (opens squares in Figures 7 and 15) fall
in the red BSS sequence.
The analysis by Szekely et al. (2007) is based on ground-based data-sets. Thus their work is limited
by the relatively poor spatial resolution 
 at least in the most central and crowded regions. 
For this reason, taking advantage of the outstanding capabilities of HST and the relatively large number 
of images acquired, we performed a variability search analysis for the selected BSS. 
Our analysis can identify only stars with a relatively short period ($\sim10$ hours), because of the
the time interval covered by our observations.
The identification of variable stars was carried out using the U, V and I time-series data separately. 
As a first step, we checked the light curves of the selected BSS visually. We considered only those stars showing
coherent evidences of variability in all the bands. With this criterion we selected nine stars, 
two of them being the two SX-Phoenicis identified by Szekely et al. (2007), the other seven being new variables. 

We analyzed the light curves of these seven candidate variable BSS by using the Graphical Analyzer of Time
Series ({\rm GRATIS}), a
private software developed at the Bologna Observatory by P. Montegriffo. {\rm GRATIS} uses both the Lomb periodogram
(Lomb 1976) and the best fit of the data with a truncated Fourier series (Barning 1963). 
The final periods adopted to
fold the light curves are those that minimize the rms scatter of the truncated Fourier series that best fit the data.
We emphasize that these are newly discovered candidate variables. No counterpart for these 
stars can be found in literature.

Four candidate new variables show in all bands a variability of $\sim0.3$ mag, a value several times larger than their typical photometric errors,
(see Figure~7). 
Periods of the order of fraction of days (between 0.15 and 0.30 days) have been estimated with GRATIS.
Such periods, coupled with the light modulation of these stars (Figure~10), would suggest that
these candidates are not pulsational variables but double systems. In particular they are likely  
WUMa stars, i.e. semi-detached binaries with ongoing mass-transfer. We named them C$_{\rm WUMa}$.
These objects are expected to be quite common within the BSS population. It is worth noting that, out of four
candidates WUMa, 
three lie along the red sequence: they are C$_{\rm WUMa}1$, C$_{\rm WUMa}2$ and C$_{\rm WUMa}4$.

The remaining three candidate variables have instead much shorter periods, of the order of P$\sim 0.04-0.07$ days, and their 
light modulation has amplitudes of about 0.05-0.1 mag. 
Their periods and light curves (shown in Figure~11) are typical of SX-Phoenicis stars. We labeled these stars as C$_{\rm SX}$. 
C$_{\rm SX}1$ and C$_{\rm SX}2$ are red BSS, while C$_{\rm SX}3$ is a blue BSS. Their position in the CMD is highlighted
by open pentagons in Figures~7 and 16. 
With the inclusion of these three variables, the number of known candidate SX-Phoenicis in NGC~362 increases from two to five.

\subsection{The BSS radial distribution}

The BSS radial distribution has been found to be a powerful tool to 
estimate the dynamical age of stellar systems (F12). In fact due to their mass (significantly larger the average) and to their relatively high luminosity,
BSS are the ideal class of object to measure the effect of dynamical processes (like dynamical friction and
mass segregation) from which the dynamical age of a stellar system can be derived. 

By combining the WFC3 and the WFI data, we have been able to study the BSS radial
distribution of NGC~362 along  its entire extension. In the
complementary WFI catalog ($r>100\arcsec$) the BSS  selection has been performed  in the (V, B-V) plane by adopting the same
magnitude limits  used for the WFC3 catalog (approximately $16.5<$V$<18.5$; see Figure~12).
We thus counted 28 BSS within the tidal radius ($r_t\sim720\arcsec$; see
Section~3). The photometric accuracy and field star contamination of the WFI catalog 
do not allow us to investigate the presence of a double sequence in the cluster
peripheries. Therefore,  we will consider BSS as a single population in the following
analysis. \\
In order to have a reference population to study the BSS radial
distribution, we selected bona-fide SGB stars in the magnitude  interval $18<$V$<18.5$,
as done for the WFC3 catalog. We have already  shown, by means  of a cumulative radial
distribution (Figure~9), that BSS in the WFC3 FoV are  more centrally concentrated than
the reference stars. Now we extend the analysis out to $r_t$ and we use the specific 
frequency $N_{\rm BSS}/N_{\rm SGB}$. We divided the FoV in five concentric annuli centered on
$C_{\rm grav}$ and in each  of them we counted the number of BSS and that of the reference population
stars. We estimated the impact of  contamination on the derived number counts by
counting the number of stars falling in the BSS and SGB selection boxes in a reference field beyond the cluster limit,
at $r>900\arcsec$ from
$C_{\rm grav}$ (right panel of Figure~12). The resulting density of contaminating field stars for 
 both the BSS and the SGB populations is $\rho\sim0.03$ stars/arcmin$^2$. We
then used this value to correct  the number counts previously obtained in the five
radial bins.  As shown in Figure~13, the radial distribution of $N_{\rm BSS}/N_{\rm SGB}$  is
monotonic: highly peaked in the innermost region and rapidly decreasing outward. \\ 
We
also computed the double normalized ratio ($R_{\rm pop}$) for BSS and  SGB, as defined in
Ferraro et al. (1993). For each radial bin, the sampled luminosities have been 
calculated from the density profile (Section~3). For the central region ($r<3\arcsec$) 
we used the best-fit slope shown in Figure~4, instead of a flat core. The value of 
$R_{\rm SGB}$ is essentially constant with a value close to the unity  (see lower panel of
Figure~13), as expected for post-MS stars (Renzini \& Buzzoni 1986). 
  Conversely,
the central value of 
$R_{\rm BSS}$ is $\sim2$, indicating that in the cluster core 
  BSS are  twice more abundant  than the reference population (which scales 
as the cluster sampled luminosity). Moreover, 
$R_{\rm BSS}$ monotonically declines for increasing distance from the center, reaching values close to
$\sim0.4$ in the most external annulus,  with no evidence of any external rising branch.\\ 
In the context of the {\it "dynamical clock"} discussed in F12, the observed 
trend of $R_{\rm BSS}$  clearly places NGC~362 in the group of dynamically old clusters ({\it Family III}).
Indeed, the comparison shown in 
Figure~14 fully confirms that the radial behavior of $R_{\rm BSS}$  is very similar to that of
other clusters  (M~75, M~80, M~79 and M~30) showing   the highest level of dynamical
evolution, with even the most external BSS already sunk toward the cluster center.  It
is worth noting that  M~30, which is the only PCC cluster in the F12 sample, belongs
to this group.  \\
Additional clues about relative dynamical-age differences within the clusters belonging 
to this family can be obtained from the detailed comparison of the BSS radial distribution shape.
In fact,  because of
its relatively  flat decreasing branch, NGC~362  appears to be similar to M~75 and M~79,
which are the youngest systems in this family. Moreover NGC~362 shows the smallest 
peak value of $R_{\rm BSS}$ 
within  {\it Family III}, again suggesting that it 
is among the dynamically youngest clusters within this group. \\
In order to quantify this impression, we determined the position of NGC~362 
 in the "{\it dynamical clock}" plane (see Figure~4 in F12), where  
the position of the minimum of   the BSS radial distribution in units of the cluster core radius, 
$\log(r_{\rm min}/r_{\rm c})$, is plotted as a function of the core relaxation time in units of the Hubble time, 
$\log(t_{\rm rc}/t_{\rm H})$. As discussed in F12, 
$\log(r_{\rm min}/r_{\rm c})$ is  the time-hand   of the dynamical clock. For clusters in {\it Family III}, 
  where no minimum can be detected, F12 assumed as $r_{\rm min}$ the 
  radius of the most distant bin in the observed BSS radial
distribution.
In the case of NGC~362, this distance corresponds to  $\log(r_{min}/r_c)\sim1.6$.
The core relaxation-time can be derived by using   equation (10) in
Djorgovski (1993) and by assuming the structural parameters obtained in  Section~3, the distance
modulus and reddening adopted in Section~5.1, and a mass-to-light ratio $M/L_V=3$ (see
Dalessandro et al. 2013 for more details).  We thus obtained that the central relaxation
time, once normalized to the Universe age ($t_H=13.7$ Gyr) is $\log(t_{rc}/t_H)\sim-2$.
Figure 15 shows the position of NGC~362 in the {\it dynamical clock plane}. 
As expected, the cluster lies close to M~75 and M~79 (showing  a strikingly similar  BSS radial distribution) and it is among 
the youngest systems of this family.

\section{Discussion}  

The accurate HST WFC3 photometry presented in this paper has revealed
the presence of two almost parallel BSS sequences in the core of
NGC~362.  This represents the second case, after M~30 (F09), for which
a double BSS sequence has been observed. The red and blue BSS
populations are well separated in the (V, V-I) CMD by
$\Delta$V$\sim0.4$ and $\Delta$(V-I)$\sim0.15$, and they nicely
overlap with the distribution of BSS shown in F09 (see Figures 7, 8).
As in the case of M~30, the red population is significantly more
centrally concentrated than the blue one (Figure 9) and their sizes
are very similar. Also the total number of BSS normalized to the total
cluster luminosity is basically the same in these two systems. In fact
we count, within $r_t$ and after field stars subtraction, 77 BSS in NGC~362 
which has $L_V\sim2\times10^5 L_{\odot}$,
and 51 in M~30 (F09) which has instead $L_V\sim 10^5 L_{\odot}$. F09
argued that blue BSS are likely the result of collisions while red BSS
are binary systems in an active phase of mass-transfer. Observational
hints supporting this interpretative scenario have been recently shown
by Lovisi et al. (2013).  A similar approach can be followed to
interpret the two sequences in NGC~362. The results are shown in
Figure 16.  Indeed the position of the blue sequence in the CMD can be
nicely reproduced by a collisional isochrone (Sills et al. 2009) of
proper metallicity ([Fe/H]=-1.31) and age $t=0.2$~Gyr and by assuming
a distance modulus $(m-M)_0=14.68$ and reddening $E(B-V)=0.05$
(Ferraro et al. 1999b). Additional support to the collisional origin
of the blue sequence can be obtained from star counts.  In fact the
number of expected BSS produced by collisions can be estimated from
equation (4) in Davies, Piotto \& de Angeli 2004 (see also Leonard et
al. 1989):
\begin{equation}
N_{COL-BSS}=0.03225~\frac{f^2_{mms} N_c n_{c,5} r_{col} m_{BSS}}{V_{rel}}
\end{equation}  
where $f_{mms}$ is the fraction of massive MS stars (i.e. stars able
to form a BSS when they collide) in the core (we assumed here
$f_{mms}=0.25$; Davies \& Benz 1995), $N_c$ is the total number of
stars in the core and $n_{c,5}$ is the density of stars expressed in
units of $10^5$ stars pc$^{-3}$ (we adopted $n_{c,5}\sim2$; Harris
1996), $m_{BSS}$ is the typical BSS mass (we used $m_{BSS}=1.2
M_{\odot}$), $r_{col}$ is the minimum separation of the two colliding
stars in solar units (we adopted $r_{col}=2 R_{\odot}$), and $V_{rel}$ is
the relative incoming velocity of binaries at infinity (we adopted
$V_{rel}=\sigma\sqrt(2)$ from Leonard et al. 1989 and the central
velocity dispersion $\sigma$ from Harris 1996).  By using equation (1)
and the quoted assumptions, we obtained that about 25 BSS are
expected to be formed in the last 0.2~Gyr by collisions.  This is in
nice agreement with the observed number of BSS (30) on the blue
sequence. While the blue BSS are all observed in the WFC3 FOV, 
equation (1) in principle concerns the entire cluster, since it is based on the assumption that COL-BSS are 
formed in the core and then spread out because of dynamical interactions. 
However, it is more likely to observe COL-BSS in the inner regions of stellar systems,
than in the outskirts. This is also confirmed by the findings of Mapelli et al. (2004, 2006) 
showing that most of the COL-BSS kicked out from the core either leave the clusters,
or sink back rapidly in the core. 
Hence, the fraction of COL-BSS outside the WFC3 FOV should be negligible.

In F09 the position of the red BSS sequence in the CMD has been found
to be well reproduced by the lower luminosity boundary defined by the
distribution of binary stars with ongoing mass-transfer, as found in
Monte Carlo simulations by Tian et al. (2006).  This boundary
approximately corresponds to the locus defined by the Zero-Age-MS
(ZAMS) shifted to brighter magnitudes by $0.75$~mag. The locus
obtained for NGC~362 is shown as a red dashed line in Figure 16. As
can be seen, red BSS lie in a sparse area adjacent to the lower
boundary in a region that we can call the {\it MT-BSS domain}
(highlighted in grey in Figure 16).  It is particularly interesting to
note that 3, out of the 4 WUMa candidates discovered in this work, lie
along the red sequence where MT-BSS are expected.

In F09, the presence of two distinct sequences of BSS has been
connected to the dynamical state of M~30, in particular to the fact
that this cluster might have recently ($1-2$ Gyr ago) experienced the
collapse of the core.  As discussed in Section 3, the dynamical state
of NGC~362 is quite debated, and controversial results are found in
the literature (Fischer et al. 1993; Trager et al. 1995; McLaughlin \&
van der Marel 2005).  The density profile cusp ($\alpha\sim-0.2$)
discussed in Section 3 is shallower than typically observed in PCC
clusters and could indicate that NGC~362 is on the verge or is
currently experiencing the collapse of the core (Vesperini \& Trenti
2010).  The advanced dynamical age of NGC~362 is also suggested by its
monotonic BSS radial distribution.  In fact, in the "{\it dynamical
  clock}" classification (F12), NGC~362 belongs (with M~30) to the
family of the highly dynamically-evolved clusters ({\it Family III}).

On the basis of this observational evidence we can argue that also in
the case of NGC~362 the presence of a double BSS sequence could be
connected to the advanced dynamical state of the cluster.  As in the
case of M~30, the fact that we observe two distinct sequences, and in
particular a well defined blue one, implies that the event that
triggered the formation of the double sequence is recent and
short-lived. If this event is connected with the dynamical evolution
of the system, it could likely be the collapse of the core (or its
initial phase). Indeed, during the collapse, the central density
rapidly increases, also enhancing the probability of gravitational
encounters (Meylan \& Heggie 1997): thus, blue BSS could be formed by
direct collisions boosted by the high densities reached in the core,
while the red BSS population could have been incremented by binary
systems brought to the mass-transfer regime by hardening processes
induced by gravitational encounters (McMillan, Hut \& Makino 1990;
Hurley et al. 2008). 

Quite interestingly, the red and the blue BSS show different radial
distributions in both M~30 and NGC~362. The origin of this feature
still remains not completely clear.  If the collapse of the core
played a role in the origin of the two sequences, then it could also
have had an impact on setting their radial distributions.  While
significant recoils are expected both for collisional products and for
hardened binaries, the fact that blue BSS are more sparsely
distributed might indicate that gravitational kicks are stronger in
the former case. Alternatively, most of the observed red BSS sank into
the cluster center because of dynamical friction and did not suffer
significant hardening during the core collapse phase. As a
consequence, they did not experienced significant recoil and they
appear more centrally segregated than blue BSS (which have been,
instead, kicked outwards during collisional interactions).  Within
such a scenario the properties of the blue BSS suggest that the core
collapse occurred very recently ($\sim0.2$ Gyr ago) and over a quite
short time scale, of the order of the current core relaxation time
($\sim 10^8$ yr; Harris 1996).\\ 
Indeed detailed spectroscopic
investigations and accurate dynamical simulations are urged to shed
light on both the nature of the red and blue BSS sub-populations and
their dynamical properties.

\acknowledgements This research is part of the project {\it
  COSMIC-LAB} (http://www.comic-lab.eu) funded by the {\it European
  Research Council} (under contract ERC-2010-AdG-267675).

\newpage

\begin{table}[!h]
\begin{center}
\begin{tabular}{|c|c|c|c|c|c|}
\hline
\hline
           &                &            &            &\\
$r_{int}$  &    $r_{ext}$  & $N_{BSS}$  & $N_{SGB}$   & $L_{samp}^{ann}$/$L_{samp}^{tot}$   \\
           &                &            &            &   \\
\hline
\hline
    $0\arcsec$  &  $13\arcsec$    &    20    &   245  & 0.13 \\
    $13\arcsec$  & $50\arcsec$	&      37    & 	1066  & 0.39 \\
   $50\arcsec$ & $100\arcsec$	&       8    & 	 423  & 0.16 \\
   $100\arcsec$  & $250\arcsec$	&       9     &  473  & 0.24 \\
  $250\arcsec$  & $720\arcsec$	&       3     &  191  & 0.08  \\
\hline
\hline
\end{tabular}
\end{center}
\caption{Number of BSS and SGB stars after field stars subtraction and fractions of sampled light in the five radial annuli used to study the BSS radial
distribution (Section 5.3)}
\label{}
\end{table}

\begin{figure}
\includegraphics[scale=0.7]{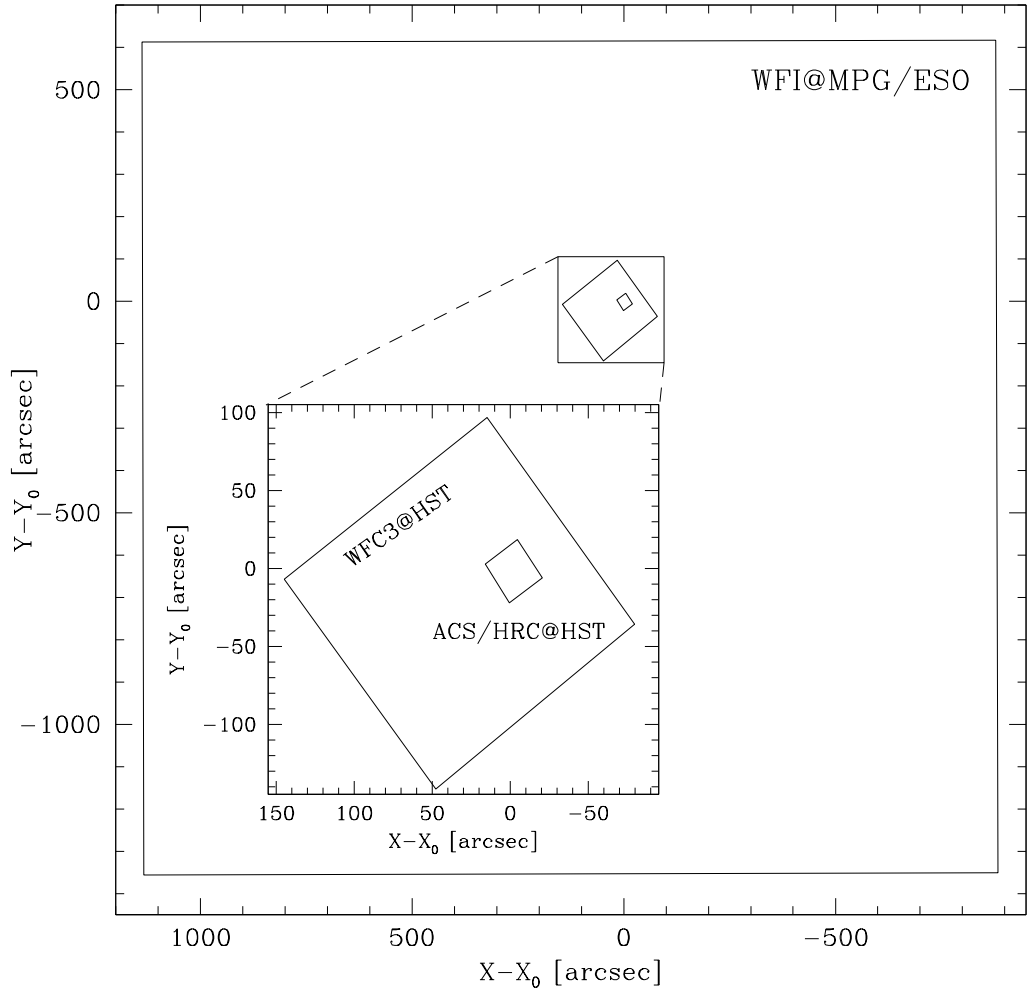}
\caption{Schematic map of the entire data-base used in this work with respect the position of $C_{grav}$ ($X_0$, $Y_0$).
The inset shows a zoomed view of the two HST sets of images. North is up, East left.}
\end{figure}

\begin{figure}
\includegraphics[scale=0.7]{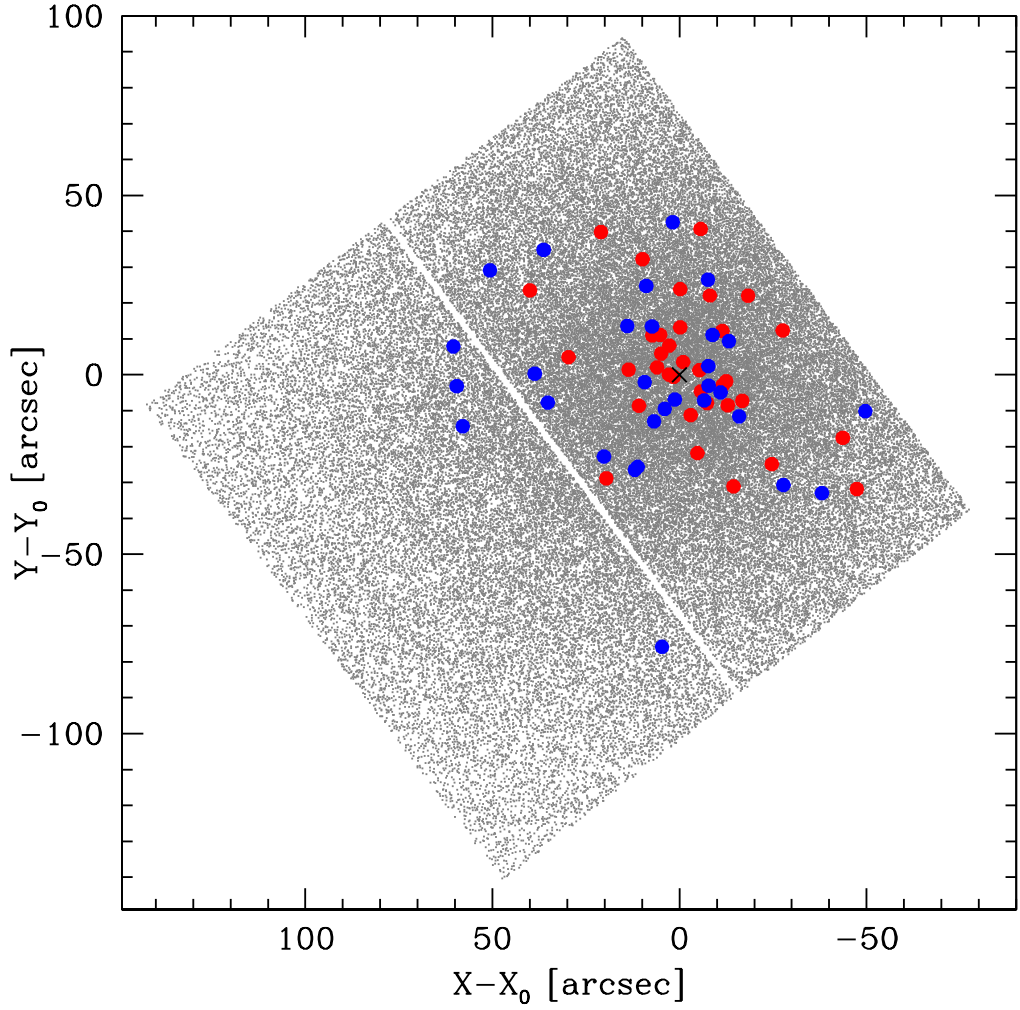}
\caption{Map of the HST WFC3 data-set with the position of the red and blue BSS highlighted.
The black cross marks the position of $C_{\rm grav}$. As in Figure~1, North is up, East left.}
\end{figure}

\begin{figure}
\includegraphics[scale=0.7]{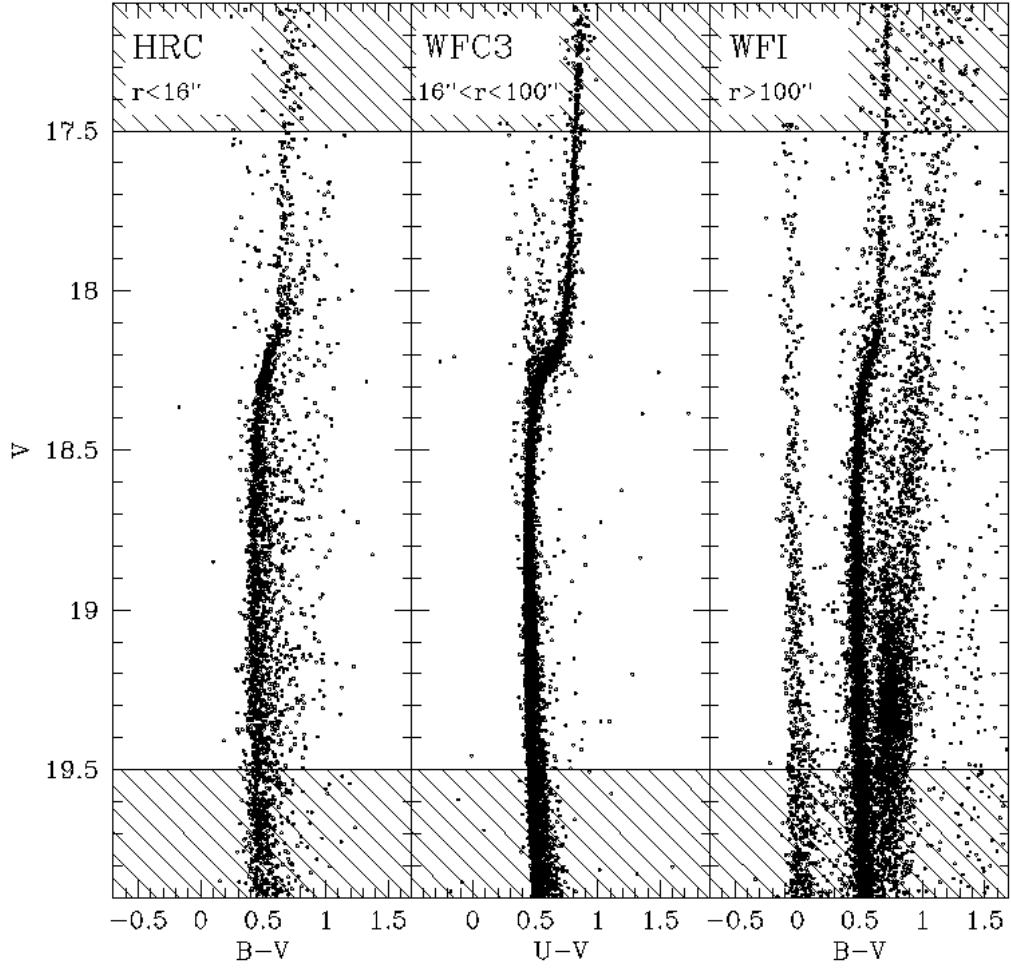}
\caption{CMDs of the three photometric samples used for the  determination of the density profile, the shaded regions delimit
 the stars included in the analysis (see Section 3).}
\end{figure}

\begin{figure}
\includegraphics[scale=0.7]{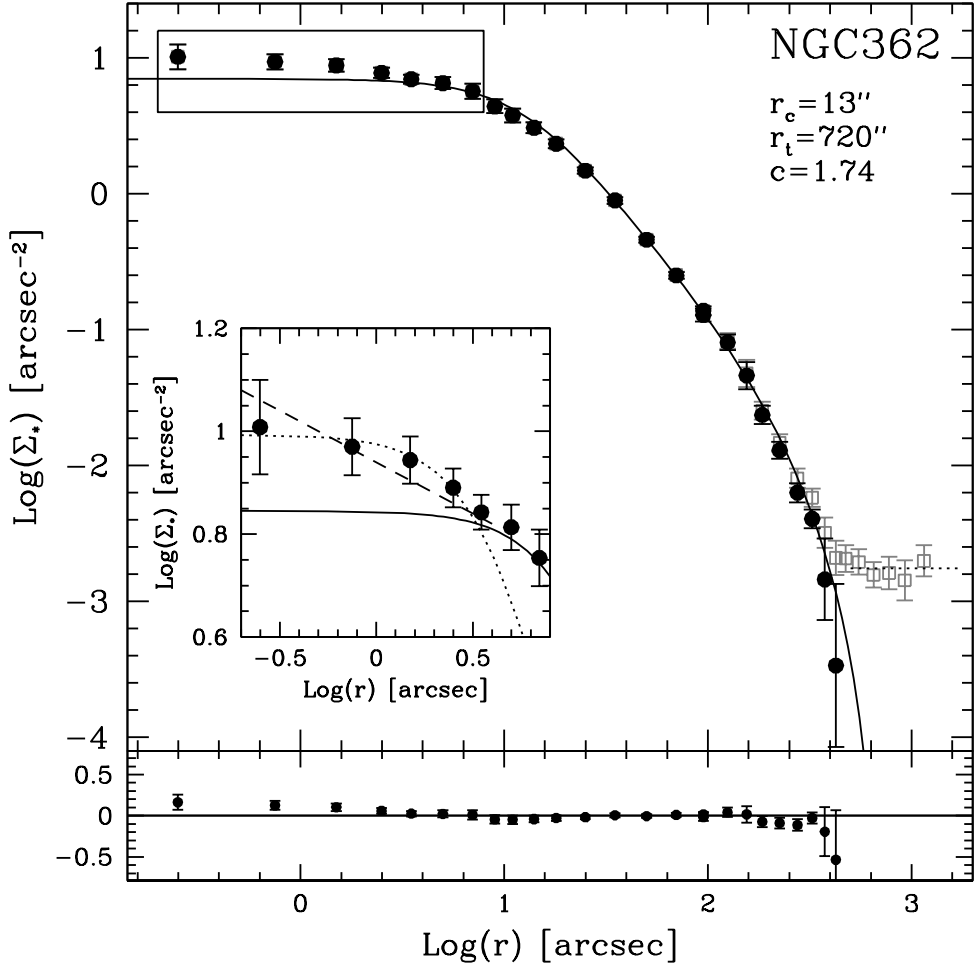}
\caption{Observed star count density profile as a function of radius (open grey squares). 
The dotted line represents the density value of the background, obtained averaging the four 
outermost points.
Black filled dots are densities obtained after background subtraction (see Section~3).
The best mono-mass King model is also over-plotted to the observations (solid line). The structural parameters are 
labelled. The lower panel shows the residuals between the observations and the best-fit model.
The inset shows a zoom on the innermost portion of the profile. The dashed line marks the power-law fitting the central cusp,
while the dotted line represents the additional King model used to reproduce the innermost $\sim3\arcsec$.}
\end{figure} 

\begin{figure}
\includegraphics[scale=0.7]{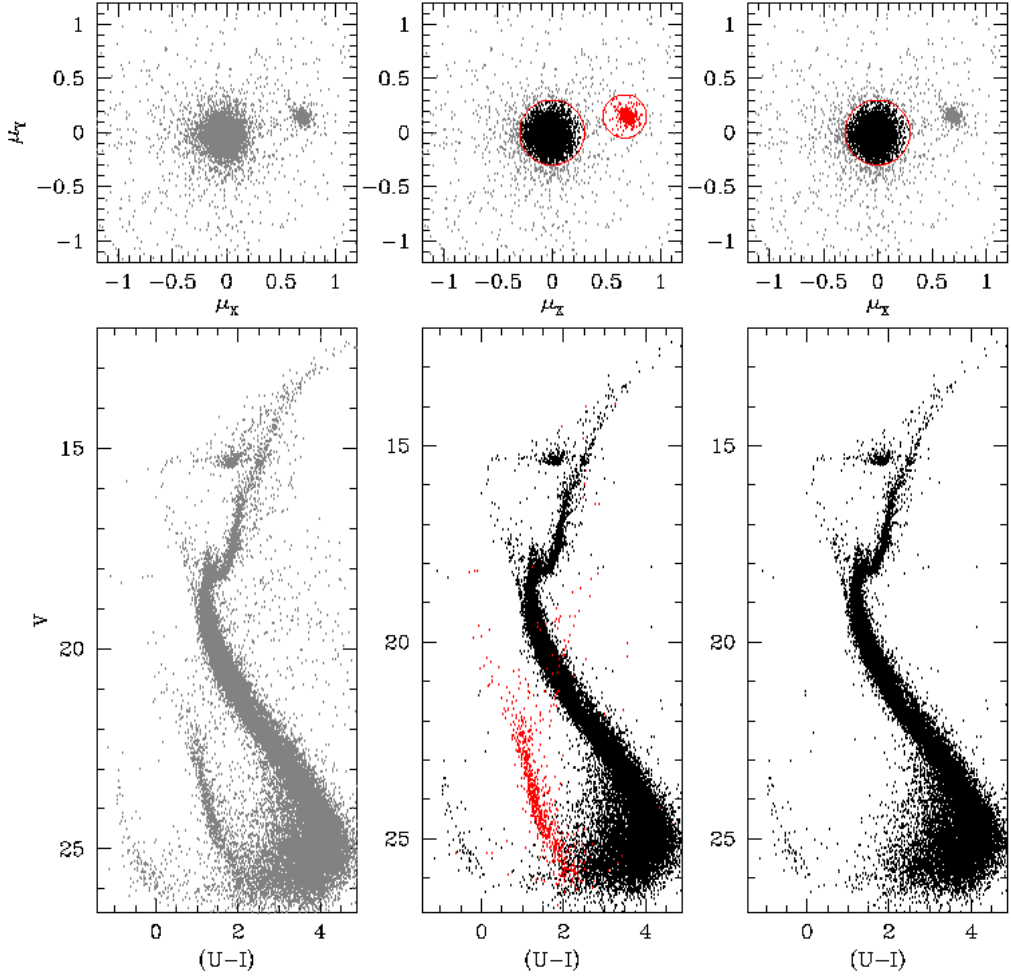}
\caption{{\it Upper panels}. In the leftmost panel VPDs in pixel/yr of all the stars identified in common between the first 
and second epoch (see Section~4). In the middle panel the cluster and the SMC populations are selected
in black and red respectively. In the rightmost panel only cluster member are selected. 
{\it Lower panels}. From left to right, (V,U-I) CMDs for all the detected stars, for stars with a  high cluster membership probability and SMC selected stars (red dots), and for cluster members  only.}
\end{figure}

\begin{figure}
\includegraphics[scale=0.7]{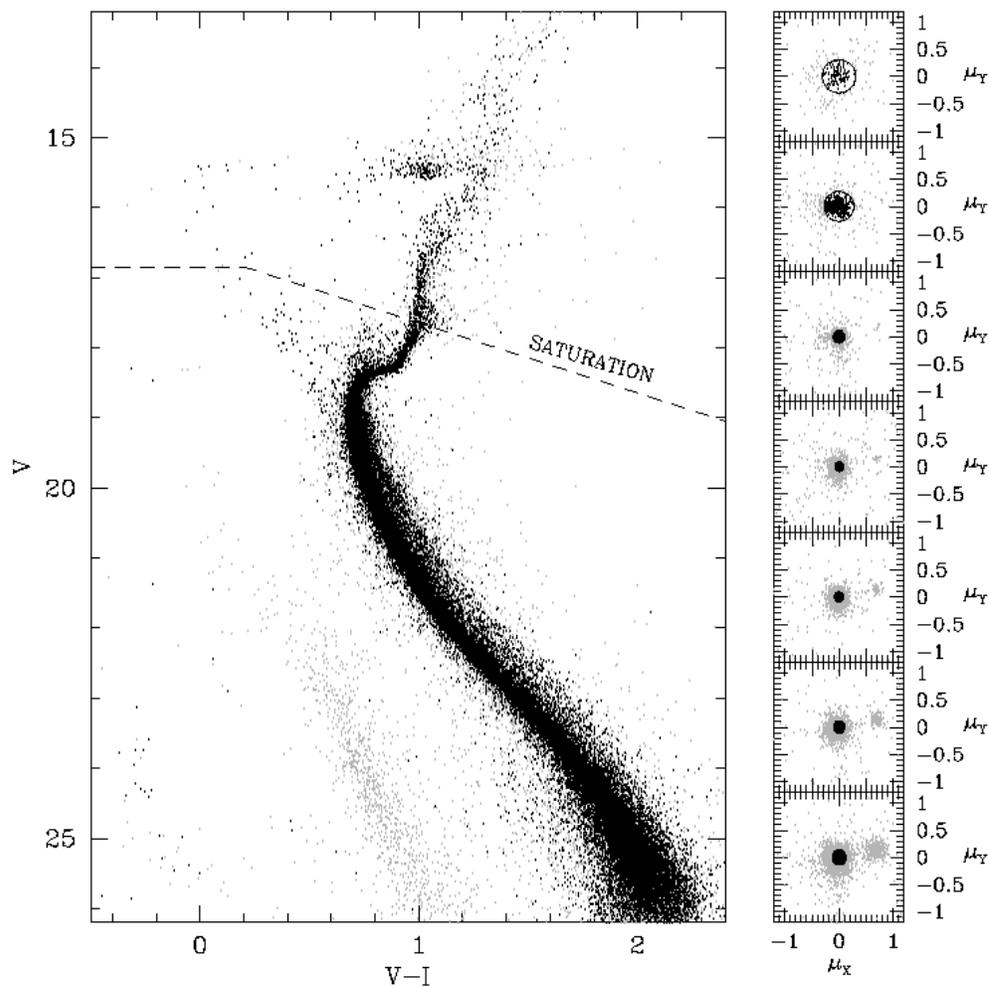}
\caption{On the left (V,V-I) CMD of stars in the WFC3 FoV. In black stars selected 
according to their membership probability, in grey stars excluded on the basis of the criteria 
highlighted in the VPDs in the right panels. On the right, VPDs at different magnitude levels. 
The distribution of stars gets broader moving to very faint and bright magnitudes because of the uncertainties on the centroid
determination. The black circle represent the $2\sigma$ fiducial region used to clean our sample from non-member stars.}
\end{figure}

\begin{figure}
\includegraphics[scale=0.7]{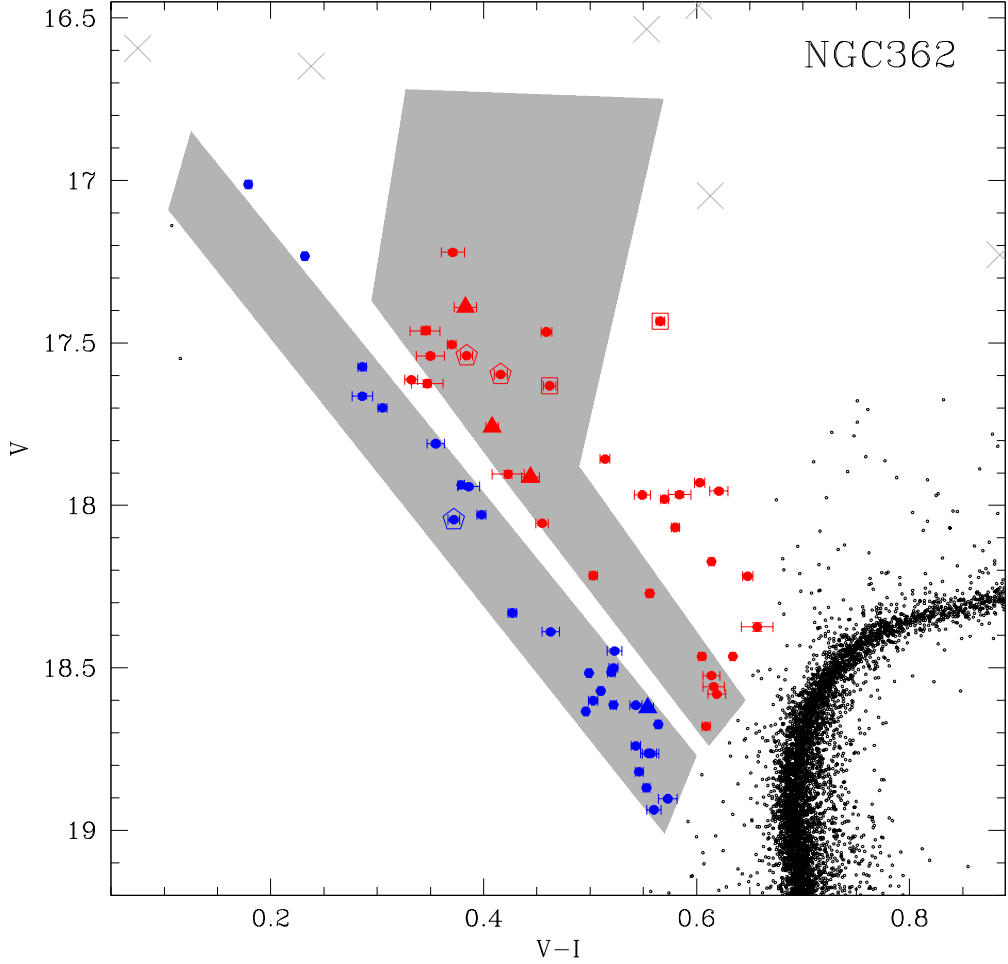}
\caption{A zoomed view of the (V,V-I) CMD of NGC~362 on the BSS region.
BSS are highlighted as red and blue symbols, and the photometric errors are shown as error bars.
The grey areas are represent the fiducial loci of the red BSS  and blue BSS 
of M~30 (from F09). Open squares are SX-Phoenicis found by Szekely et al. (2007).
Open pentagons and filled triangles are respectively SX-Phoenicis and WUMa stars identified in this work.
Grey crosses are stars excluded for saturation or non linearity problems.
 }
\end{figure}

\begin{figure}
\includegraphics[scale=0.7]{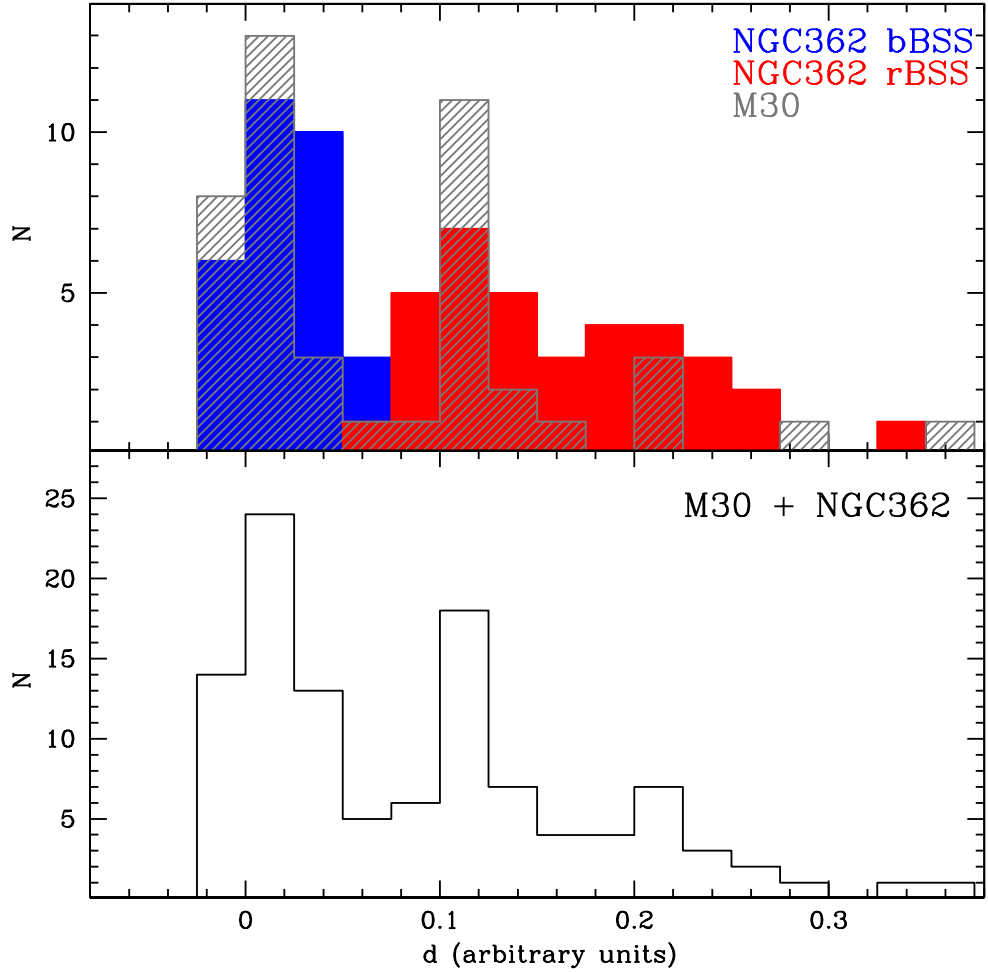}
\caption{{\it Upper panel}. Histogram of the distances (d) of red BSS  (red area) and blue BSS (blue area) 
of NGC~362 in the (V,V-I) CMD from the same reference line used in F09 and passing through the blue sequence. For comparison we superimposed the
distance distribution of the BSS in M~30 (grey shaded histogram). {\it Lower panel}. Histogram of distances for
the combined sample of BSS observed in M~30 and NGC~362.}
\end{figure}

\begin{figure}
\includegraphics[scale=0.7]{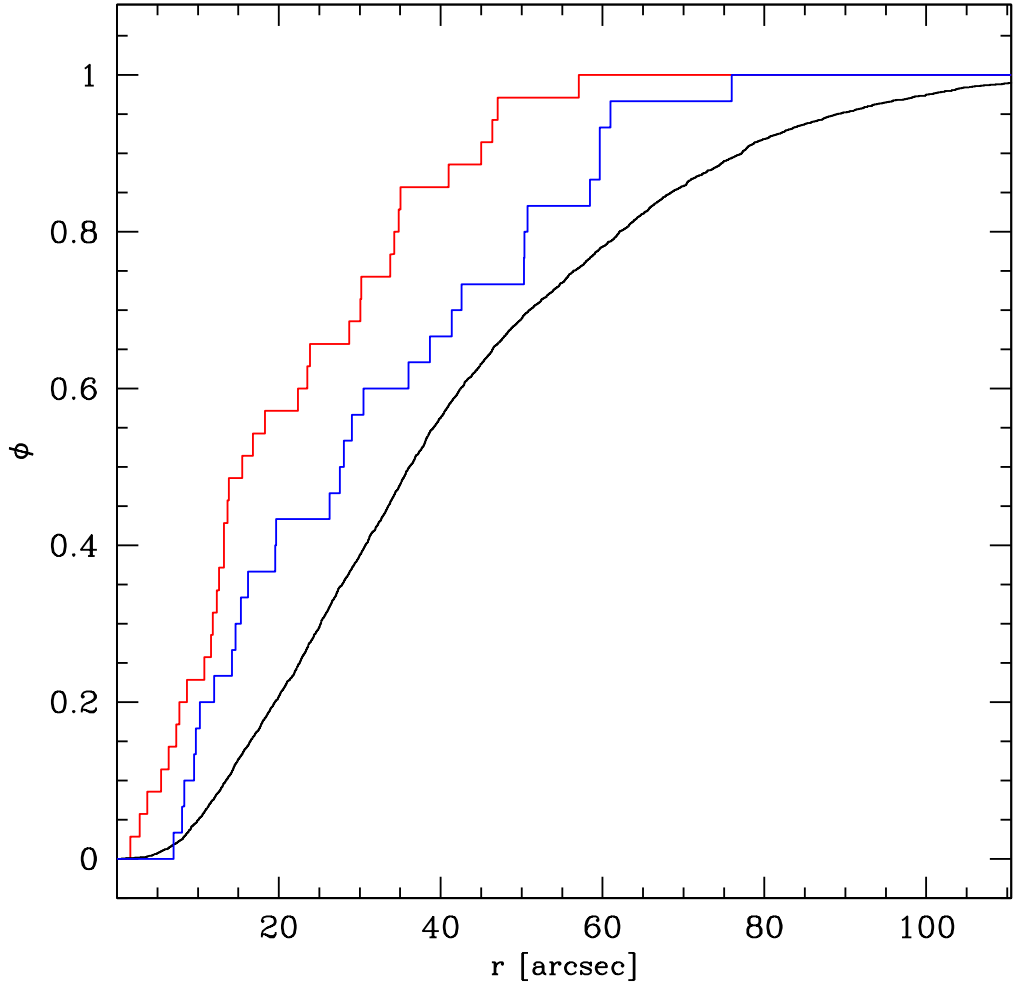}
\caption{Cumulative radial distribution of the red and blue BSS samples. 
In black  the distribution of 
SGB stars, taken as reference population. This analysis is limited to the WFC3 FoV which extends 
to a distance from $C_{grav}$ $r\sim100\arcsec$, corresponding to about $8-9 r_c $.}
\end{figure}

\begin{figure}
\includegraphics[scale=0.7]{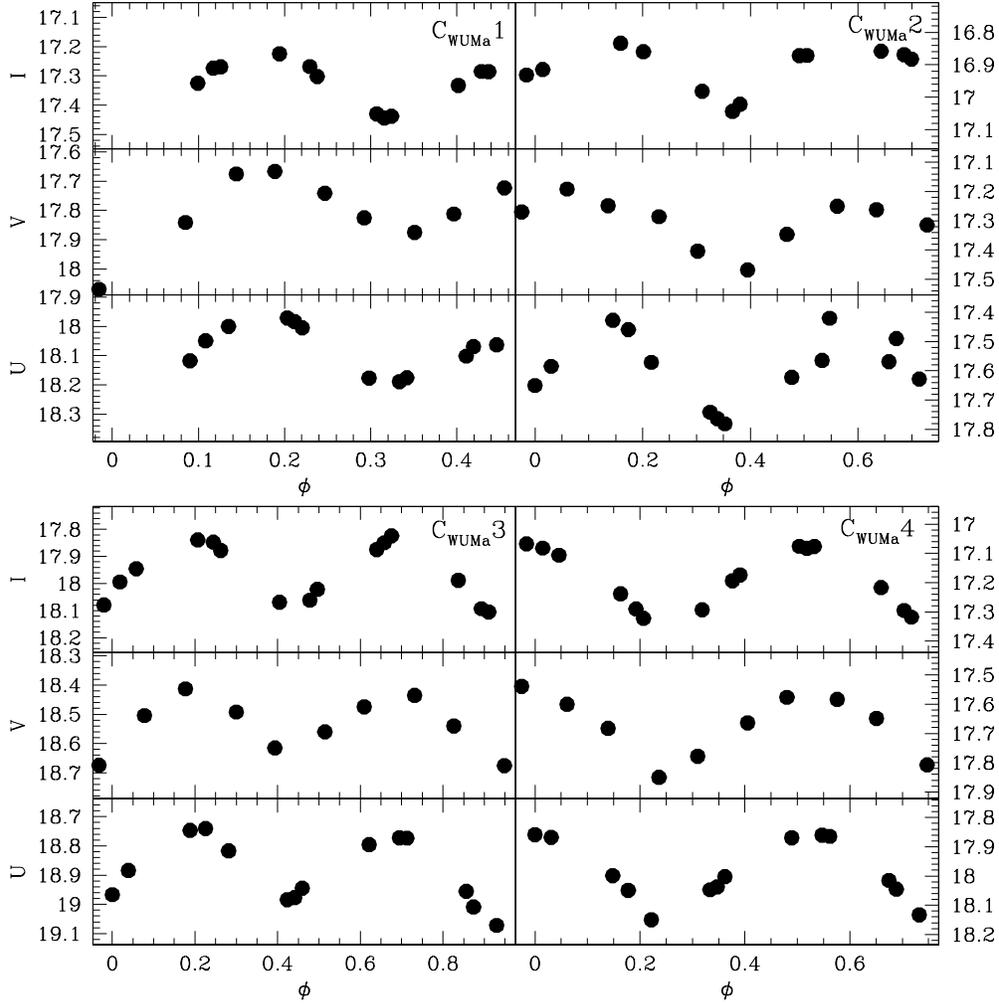}
\caption{Light curves folded to the best period solutions (obtained as described in Section~5.2)
of the four candidates WUMa $C_{WUMa}$ identified in this work.}
\end{figure}

\begin{figure}
\includegraphics[scale=0.7]{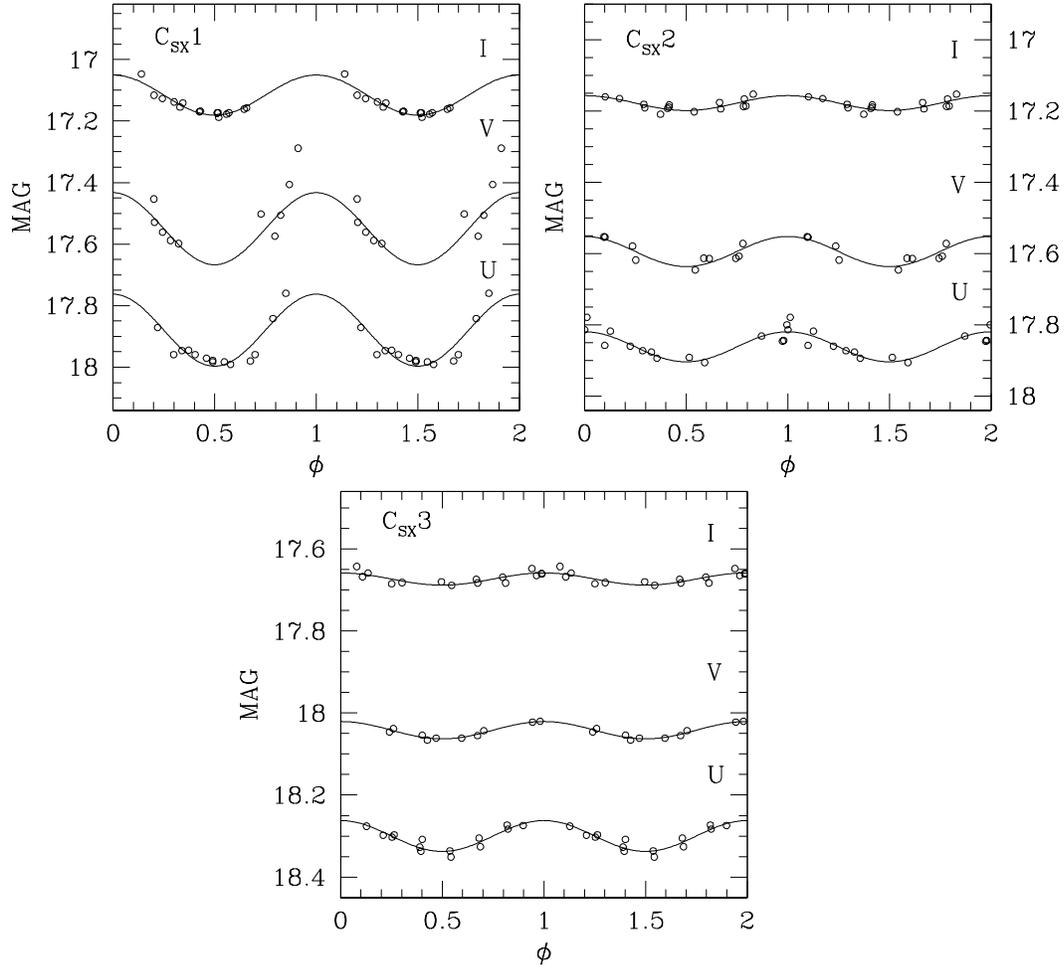}
\caption{As for Figure~10 for the three candidate SX-Phoenicis identified in this work.
The black lines are the best-fit obtained for each light curve in each band with GRATIS.}
\end{figure}

\begin{figure}
\includegraphics[scale=0.7]{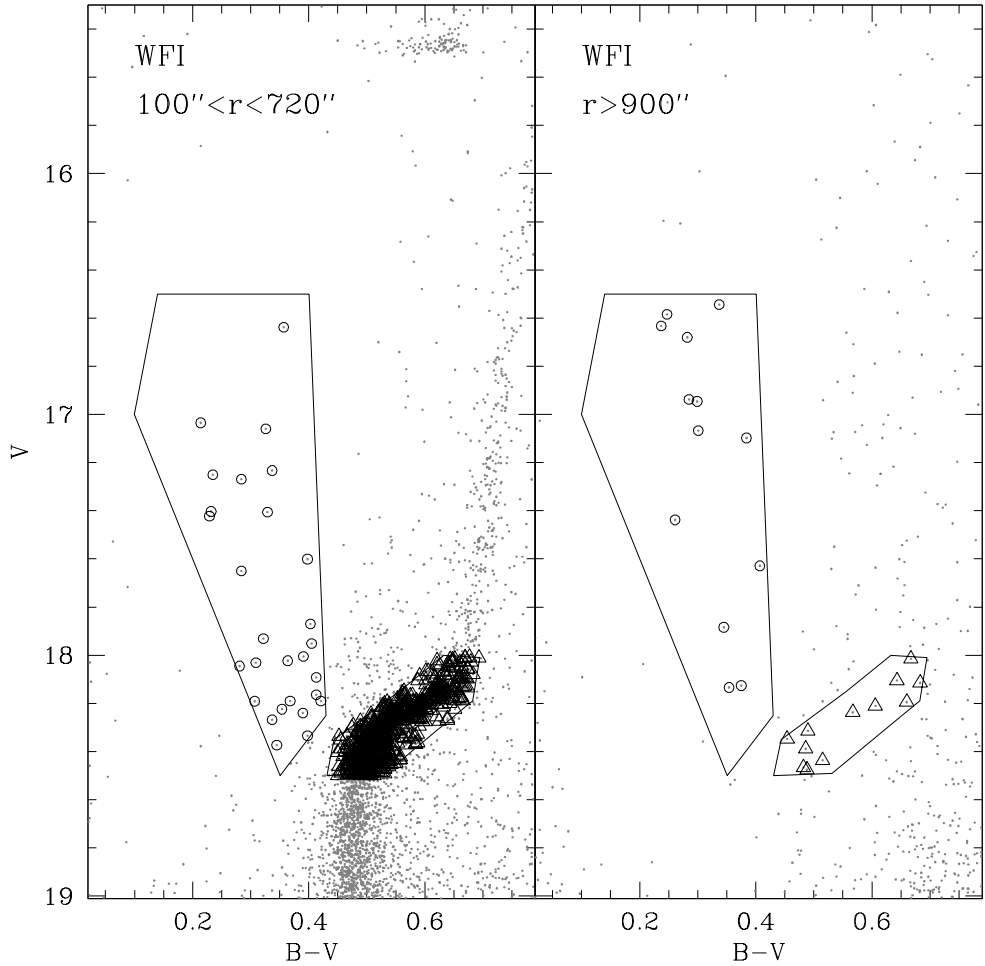}
\caption{{\it Left panel:} (V, B-V) CMD of the WFI sample for the stars with $100\arcsec<r<720\arcsec$.
Open circles are the selected BSS, open triangle are selected SGB stars. {\it Right panel} WFI CMD of the
control field at $r>900\arcsec$. The objects lying with the BSS and SGB selection boxes have been counted and used  to estimate the Galactic field contamination.}
\end{figure}

\begin{figure}
\includegraphics[scale=0.7]{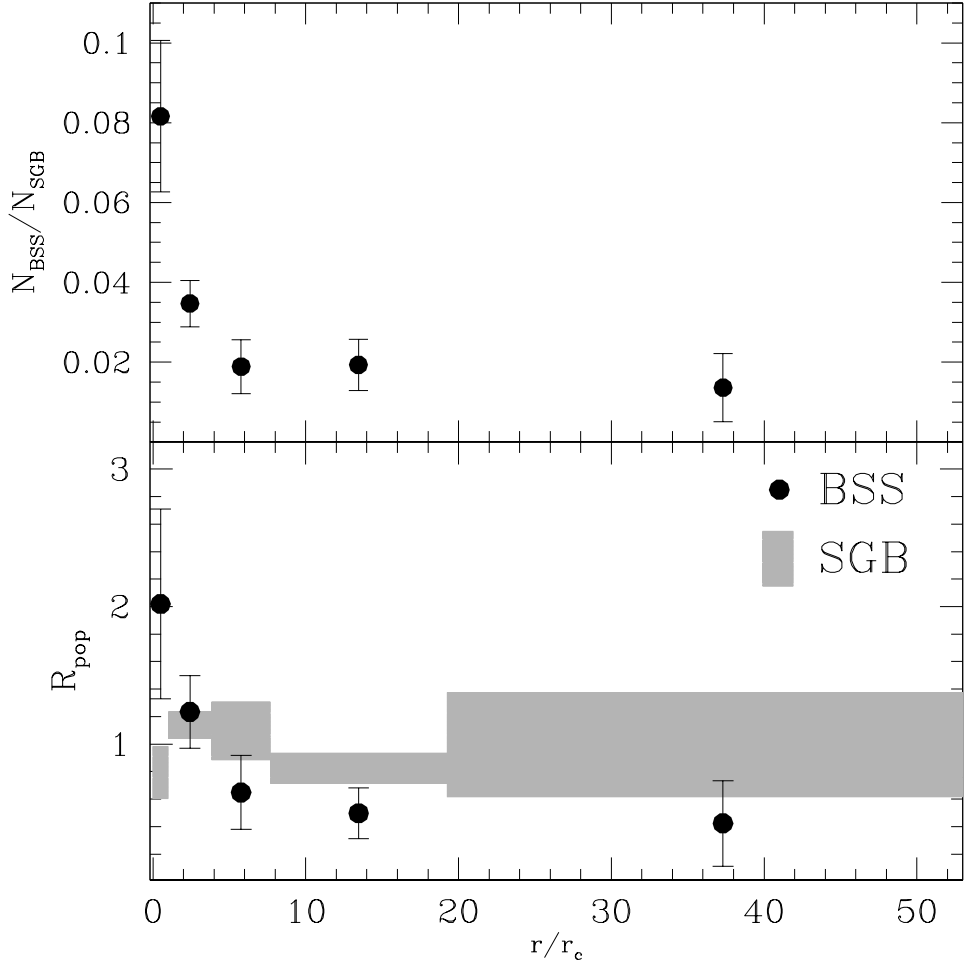}
\caption{{\it Upper panel:} radial distribution of the specific frequency $N_{BSS}/N_{SGB}$  as a function of the distance 
from $C_{grav}$ normalized to $r_c$. {\it Lower panel:} radial distribution of the double normalized ratio of BSS and SGB.}
\end{figure}

\begin{figure}
\includegraphics[scale=0.7]{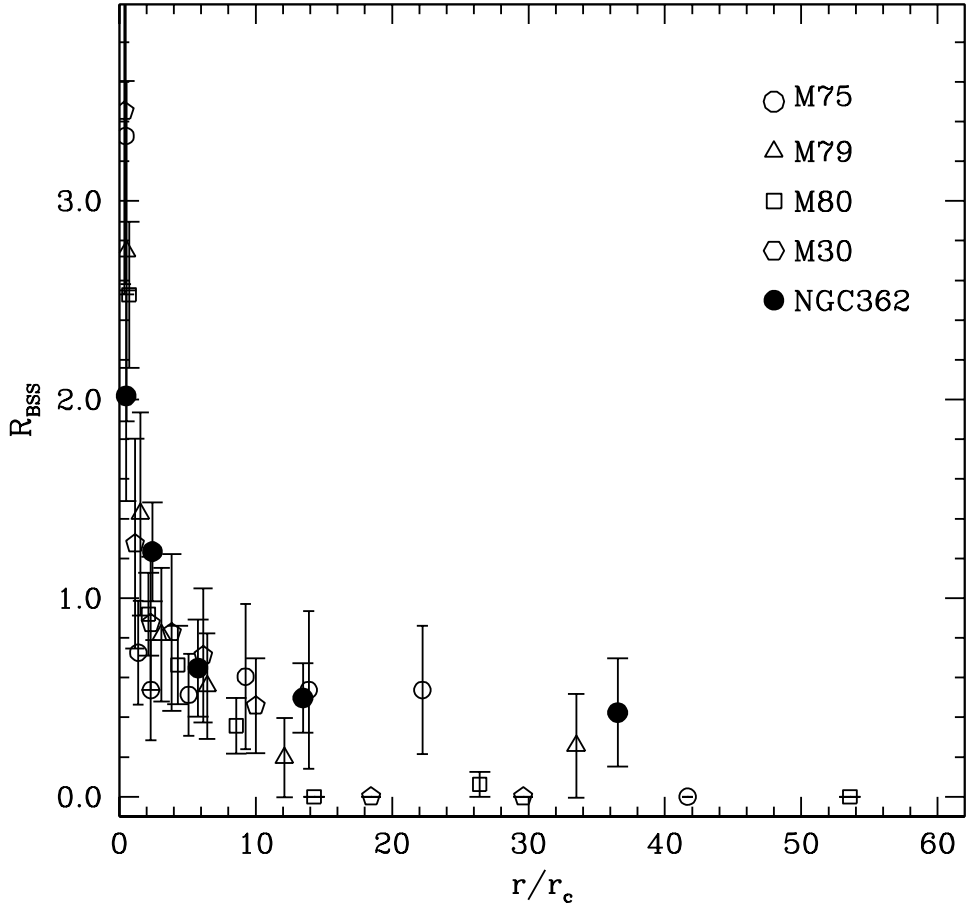}
\caption{Radial distribution of the BSS double normalized ratio of NGC~362 compared to that of the other clusters grouped in Family III by Ferraro et al.
(2012). 
The distribution in NGC~362 well resembles that of  M~75 and M~79, decreasing less rapidly than
that of other (more evolved) clusters in this group. NGC~362 also shows the smallest central peak value ($R_{BSS}\sim 2$) within in this family.}
\end{figure}

\begin{figure}
\includegraphics[scale=0.7]{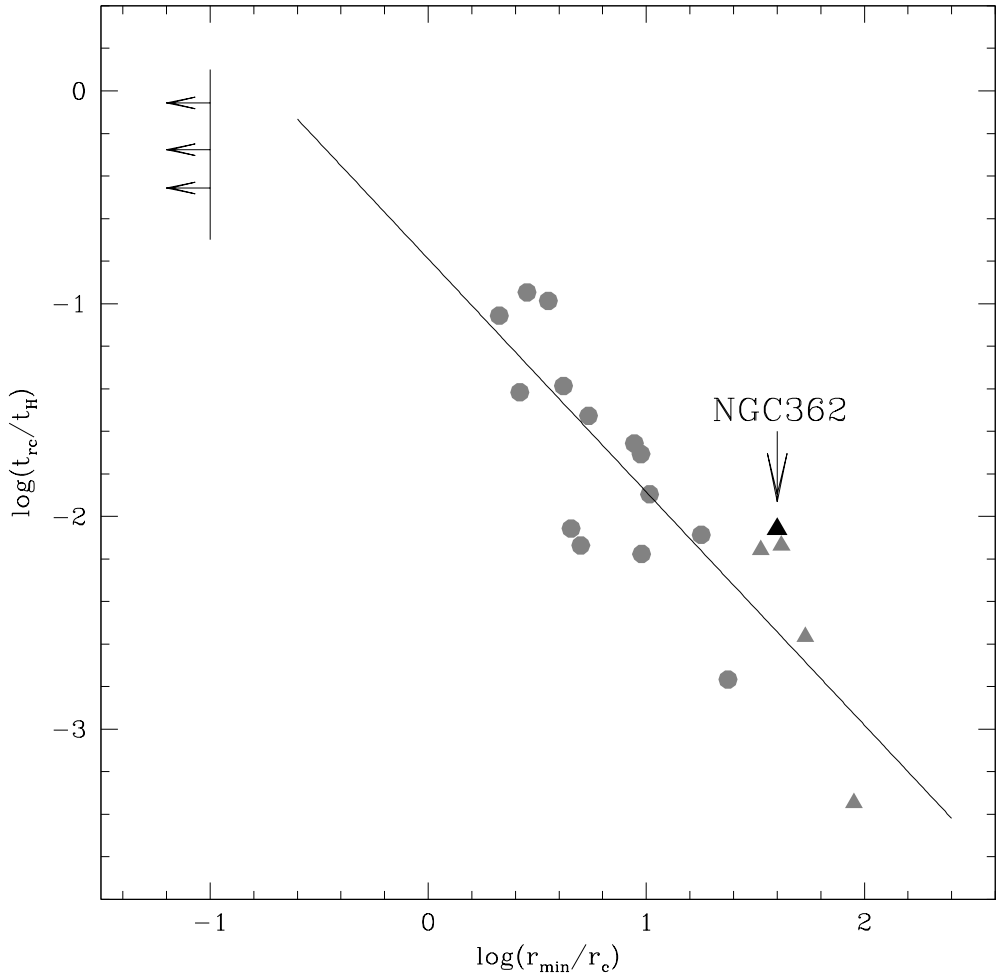}
\caption{"Dynamical clock" plot: relaxation time at the cluster center normalized to the age of the Universe $t_{\rm H}$ as a function of the time-hand   of the dynamical clock, 
$\log(r_{\rm min}/r_{\rm c})$. 
Black arrows are the lower limits of the minimum position for {\it Family I} clusters. 
Filled circles are {\it Family II} members, while grey triangles are {\it Family III} clusters.
 The black triangle highlights the position of NGC~362.}
\end{figure}

\begin{figure}
\includegraphics[scale=0.7]{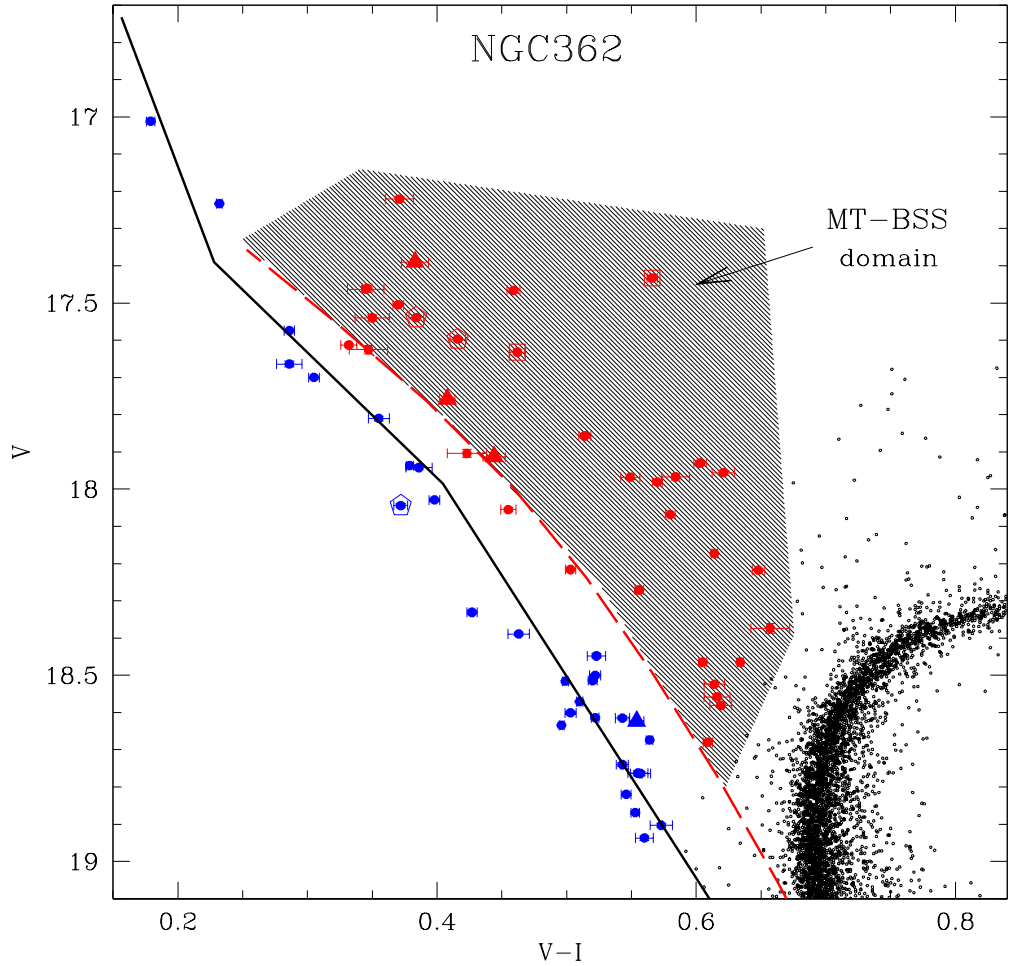}
\caption{As in Figure~7, the grey shaded area (here defined  as ``MT-BSS domain'') approximately indicates 
the region populated by mass-transfer binaries in M~67 (Tian et al. 2006),
"translated" into the CMD of NGC~362. The solid black line is a 0.2 Gyr collisional isochrone (Sills et al. 2009).}
\end{figure}

\end{document}